\documentclass[%
 reprint,
 amsmath,amssymb,
 aps,
]{revtex4-2}

\usepackage{graphicx}
\usepackage[utf8]{inputenc}
\usepackage[usenames,dvipsnames]{xcolor}
\usepackage[ruled]{algorithm2e}
\usepackage{hyperref}
\usepackage[parfill]{parskip}

\newcommand{\comment}[1]{}

\newcommand{\mps}{\mathrm{m\cdot s^{-1}}}
\newcommand{\vctor}{\vec}

\begin{document}

\title{From microscopic droplets to macroscopic crowds: Crossing the scales in models of short-range respiratory disease transmission, with application to COVID-19}

\author{Simon MENDEZ}
\email{simon.mendez@umontpellier.fr}
\affiliation{%
IMAG, Univ. Montpellier, CNRS, Montpellier, F-34095, France
}%
\author{Willy GARCIA}
\author{Alexandre NICOLAS}%
 \email{alexandre.nicolas@polytechnique.edu}
\affiliation{%
Institut Lumi\`ere Mati\`ere, CNRS, Univ. Claude Bernard Lyon 1, Villeurbanne, F-69622, France
}%

\date{\today}

\begin{abstract}
Short-range exposure to airborne virus-laden respiratory droplets is now acknowledged as an effective transmission route of respiratory diseases, as exemplified by COVID-19. 
In order to assess the risks associated with this pathway in daily-life settings involving tens to hundreds of individuals, the chasm needs to be bridged between fluid dynamical simulations of droplet propagation and population-scale epidemiological models. We achieve this by coarse-graining  microscopic droplet trajectories (simulated in various ambient flows) into spatio-temporal maps of viral concentration around the emitter and coupling these maps to field-data about pedestrian crowds in different scenarios (streets, train stations, markets, queues, and street cafés).
 At the scale of an individual pedestrian, our results highlight the paramount importance of the velocity of the ambient air flow relative to the emitter's motion. This aerodynamic effect, which disperses infectious aerosols and thus mitigates short-range transmission risks, prevails over all other environmental variables.
At the crowd's scale, the method yields a ranking of the scenarios by the risks of new infections that they present, dominated by the street cafés and then the outdoor market.
While the effect of light winds on the \emph{qualitative} ranking is fairly marginal, even the most modest ambient air flows dramatically lower the quantitative rates of new infections.
The proposed framework
was here applied with SARS-CoV-2 in mind, but its generalization to other airborne pathogens and to other (real or hypothetical) crowd arrangements is straightforward. 
\end{abstract}

\maketitle
\section{Introduction}

The theoretical combat against respiratory infections stretches over a whole gamut of lengthscales: These diseases are caused by viruses or bacteria, which measure several tens of nanometers ($10^{-8}$~m) and around one micron ($10^{-6}$~m) of diameter, respectively; these germs are (directly or indirectly) transmitted from person to person, using as carriers respiratory droplets (ranging from several hundred nanometers to several hundred microns in diameter) that are mostly expelled through the centimetric ($10^{-2}$~m) mouth gap and transported over some $10^{-1} \sim 10^{0}$ meters.
 The daunting \emph{fundamental} challenge of bridging so many scales to model the transmission of these diseases has also become 
 an imperious \emph{practical} necessity since a new coronavirus, the `severe acute respiratory syndrome coronavirus 2' (SARS-CoV-2) was identified during a first epidemic outburst in Wuhan, China, in December 2019. Since then, it has spread all over the world and caused the COVID-19 pandemic, which is officially accountable for more than 550 million cases and 6.4 million deaths as of July 2022 \cite{Johns_Hopkins_CSSE}. 

Regarding the transmission pathways of respiratory diseases, direct impact of droplets of respiratory fluids on the nasal or oral mucosa of the susceptible individual as well as contact with droplet-contaminated surfaces (the so-called fomites) have long been identified as possible routes. The susceptible individual may also get infected after inhaling pathogen-laden aerosols exhaled by a contagious person, a mechanism termed
airborne transmission \cite{Bourouiba2020fluid,Morawska2021physics}. Here, the term aerosol refers, and will henceforth refer, to all respiratory droplets small enough to dwell in the air for at least a few seconds and to be inhaled by somebody through their nose or mouth.
The prevalence of this airborne transmission route has been increasingly acknowledged \cite{Chen2020short-range,zhang2020identifying,Morawska2021physics,azimi2021mechanistic,Wang2021airborne}, especially in crowded indoor environments that led to well documented superspreading events \cite{Wang2021airborne,Morawska2020time,Miller2021transmission,Greenhalgh2021ten}. The alarm has also been raised with respect to crowded outdoor settings  \cite{WHO_transmission} (e.g., at mass sports events), where accumulation of virus-laden aerosols in the air is implausible but short-range exposure can occur; nevertheless, the actual risks that they present have been a bone of contention \cite{Bulfone2021outdoor}.

To assess how the disease may spread in crowds, modeling the emission, transport, and inhalation of respiratory droplets
is an appealing option and has been widely used in COVID-19-related studies. However, modeling transmission is a major challenge, owing to the sensitivity of droplet propagation to environmental factors such as temperature, humidity and wind \cite{Chong2021extended,Bale2022quantifying,Wang2021short-range}, as well as the uncertainty about the sizes of emitted droplets \cite{rosti2020fluid} or the person-to-person variability \cite{asadi2020efficacy}, among others. 
Moreover, \emph{microscopic} studies of droplet propagation,  supposed to describe the evolution of droplets most accurately, are generally limited to static scenarios involving two people facing each other. At the other extreme, most \emph{macroscopic} models focusing on indoor transmission assume a well-mixed environment \cite{Miller2021transmission,Bazant2021guideline,Buonanno2020estimation,Salinas2022improved} and thus overlook short-range exposure, which is the main source of risks outdoors.

Here, we endeavor to bridge the gap between detailed micro-environment studies and their macroscopic counterparts, by building on the framework outlined in \cite{garcia2021model}. In this framework, field data about pedestrian behavior (including the interpedestrian distances, interaction durations, head orientations, etc.) are coupled to concentration maps of virus-laden particles
 exhaled by a (supposedly contagious) individual in the crowd in order to assess the number of susceptible people that this individual would infect. Unfortunately, these concentrations maps were so far largely \emph{ad hoc} and rested on crude modeling assumptions. In this paper, the connection with the microscale is fully established thanks to genuine computational fluid dynamical (CFD) simulations of droplet propagation (performed using large-eddy computations to account for flow turbulence) and converted into concentrations maps via a transparent coarse-graining method. A variety of ambient conditions, notably air flow velocities, are considered, which enables us to quantify the effect of ambient air flows, the walking speed, as well as the pedestrians' activity (breathing or talking). Overall, the framework provides an unprecedented means to assess the risks of new infections \emph{via short-range} exposure in arbitrary (real or hypothetical) crowd settings. Incidentally, while we have here chosen model parameters corresponding to SARS-CoV-2, the framework can easily be generalized to any pathogen with airborne transmission.  
 
 In the next section, the scientific context of the work with regard to airborne transmission is further clarified. Next, Sec.~\ref{sec:methods} describes our methodology, from the macroscopic model to assess the risk of new infections to the microscopic simulations of droplet transport. Sec.~\ref{sec:results_micro} then exposes the risks of transmission from a single infected person exhaling in different ambient flows and for different walking speeds. Finally, Sec.~\ref{sec:results_macro} completes the connection with the macroscopic crowd by assessing the risks of new infections in real daily-life situations (on the street, at a train station, at the market, at a caf\'e), with a focus on the effect of the wind. 

\section{Scientific context of the study}

Whenever one breathes, talks, pants, coughs, or sneezes, droplets of respiratory fluids possibly containing pathogens are expelled through one's mouth and, to a much lesser extent, nose \cite{asadi2019aerosol,Bourouiba2020fluid,Morawska2021physics,Wang2021airborne}. In the case of COVID-19, the largest droplets thus produced had initially been thought to be liable for disease transmission. However, airborne transmission by inhalation of their smaller counterparts is now supported by robust evidence and was acknowledged by the World Health Organization (WHO) in Spring 2021, after months of controversy \cite{Wang2021airborne,Lewis2020coronavirus,Randall2021how}: SARS-CoV-2's ability to be transmitted via aerosols is now well established  \cite{Morawska2020time,zhang2020identifying,Miller2021transmission,Greenhalgh2021ten}. It follows that  closed, poorly ventilated spaces are particularly propitious for transmission \cite{Bhagat2020effects,Bazant2021guideline,Peng2022practical}, insofar as the smallest aerosols, of less than a few microns, can linger in the air for hours and accumulate in rooms. This opens the door  for long-range airborne transmission, which cannot be avoided by social distancing. Nevertheless, airborne transmission may also occur at short distances, when a susceptible person inhales infected aerosols close to the emitter, where they are more concentrated. Large aerosols, with diameters $d_p$ up to 100 microns according to the recent literature  \cite{Wang2021airborne,Randall2021how,Greenhalgh2021ten,fennelly2020particle}, may also be inhaled  before their sedimentation. The sedimentation speed $v_g$ in quiescent air may be estimated by balancing the gravity $g$ and drag forces at low Reynolds numbers, where the Stokes law involving the air viscosity $\eta$ holds, and neglecting the density of the air compared to that of the droplet, $\rho_p$, viz,
\begin{equation*}
v_g\approx \frac{D^2 \rho_p g}{18 \eta}.
\end{equation*}
A droplet of fixed diameter $d_p=100\,\mathrm{\mu m}$ thus sediments at a speed $v_g\approx 0.3\,\mps$ (it will thus hit the ground in 5~s it it falls from a height of 1.5~m)  \cite{Netz2020physics}. To gauge whether it can be lifted up by an inhaling flow, bear in mind that the latter has a typical speed of a few tens of centimeters per second around the nostrils  ($0.22\,\mps$ in  \cite{Murakami2004analysis}). 

To what extent is the scenario of indoor transmission altered by outdoor settings? The most obvious difference is that aerosols are dispersed outdoors, which wards off the risk of long-range airborne transmission \cite{Wang2021airborne} and ascertains the mitigation efficiency of social distancing. On the other hand, the risks due to short-range exposure persist: one may inhale the small respiratory droplets emitted by a sick person in one's immediate vicinity, the definition of which depends on the expiratory activity (for instance, uncovered sneezes propel droplets several meters ahead of the emitter \cite{Bourouiba2020fluid,Abkarian2020speech,Chong2021extended,bourouiba2014violent}). Besides, short-range exposure outdoors may differ from indoors because, all variations in temperature and humidity conditions left aside, it involves stronger wind and air flows. Note, however, that (moderate) air currents may also be worth considering indoors, where they are also present \cite{Poydenot2021risk}. In the context of the COVID-19 pandemic, transport by the wind has alternatively been thought to favor transmission by extending the spatial reach of droplets
and inhibit it by quickly dispersing pathogens \cite{Bale2022quantifying,Poydenot2021risk,Poydenot2021turbulent}. 

In practice, for prevention policies, the risks incurred in crowded indoor environments have been highlighted by famous superspreading events \cite{Wang2021airborne,Morawska2020time,Miller2021transmission,Greenhalgh2021ten}. Outdoor infections have also been documented \cite{shen2020community,leclerc2020settings,Bulfone2021outdoor}, but very generally looked down upon as secondary.  
Still, crowded outdoor settings are still listed among the risky configurations, e.g., in WHO's animation for public information   (accessed in July 2022) \cite{WHO_transmission}. 
In particular, mass outdoor events such as sports games have been suspected to promote viral spread in periods of low viral prevalence  \cite{Alfano2022covid,Cuschieri2021mass}, but
the specific contribution of outdoor transmission in these occurrences remains unclear  \cite{Bulfone2021outdoor,Sassano2020transmission,Walsh2022effectiveness}, notably because many such events mix indoor and outdoor settings  \cite{Cuschieri2021mass,EuropeanCDC2020August,Brown2021outbreak}. In addition, retrospective studies may be biased towards an overestimation of the impact of specific large events, which are more closely monitored \cite{Suner2022association,Miron2022outdoor}. Despite these difficulties, the question of the regulation of these events is vested with special interest, given their huge economic and social role; assessing the transmission risks that they present is thus of paramount importance to hit the right balance between public safety and social impact \cite{Harris2021safe}. 

To this end, some randomized controlled trials have been conducted, in particular for indoor concerts \cite{Llibre2021screening,Delaugerre2022prevention}, but general conclusions can hardly be reached from the small pool of such studies. 
Numerical studies provide a means to circumvent these limitations; indeed, their replicability enables researchers to test assumptions, investigate the effect of different parameters, relate behaviors to transmission risks and build a mechanistic picture of the risks in such contexts. The COVID-19 pandemic has prompted an unprecedented effort from the fluid mechanics community to probe the transport of respiratory droplets after their emission, in particular using CFD \cite{Abkarian2020speech,Chong2021extended,Bale2022quantifying,Cortellessa2021,Giri2022colliding,vuorinen2020modelling,Singhal2022virus}, which has shed light on the sensitivity of this propagation to the environment \cite{Chong2021extended,Bale2022quantifying,Wang2021short-range}. Simulations have thus considered diverse environmental settings, as well as diverse expiratory events, including coughing \cite{vuorinen2020modelling,Chong2021extended,mariam2021cfd},
sneezing \cite{Wang2021short-range,mariam2021cfd}, speaking \cite{Abkarian2020speech,vuorinen2020modelling,mariam2021cfd,Bale2022quantifying} and breathing \cite{Abkarian2020speech}. Coughs, in particular, have received a lot of attention, but in this paper we put the focus on talking and breathing through the mouth, because we have deemed that direct exposure to coughs (not covered by the emitter's hand and directed towards the receiver's face) is fairly rare and, in addition, talking for 1 minute produces approximately as many droplets (i.e., a few thousand altogether) as one cough \cite{dhand2020coughs}.

Risk assessment must also involve a model for inhalation.  
In simulations, specific areas (nose, mouth, eyes) that can be impacted or traversed by droplets may be marked in the simulation domain, in order to 
gauge the relative risks raised by droplet impact and inhalation \cite{Cortellessa2021,Chen2020short-range} or to quantify the protective effect of the exhalation of the susceptible person in a conversation \cite{Giri2022colliding,Singhal2022virus}, for instance.
The inhalation volume of a passive scalar may also be used to assess the risk \cite{Villafruela2016influence}.
Leaving aside inhalation, the local concentration of virus in a region of interest may be monitored, as a proxy for the infection risk \cite{Wang2021short-range}; the need to simulate the susceptible person at each position that they may occupy is thus bypassed. Time may be involved by  comparing the quantity of inhaled virus over time to an infectious dose \cite{Bale2022quantifying,Cortellessa2021,Giri2022colliding,Singhal2022virus,Yang2020towards,Bagheri2021upper} 
via  a dose-response model \cite{SzeTo2010review,Mittal2020mathematical}. This quantity can be measured in absolute terms, as a number of viral copies, which requires specifying the viral titer in the emitter's respiratory fluids and the minimal infectious dose, or in terms of quanta of infection, if the number of emitted virus is rescaled by the infectious dose \cite{Buonanno2020estimation}. 
In either case, an additional step is required to bridge the gap between such studies of very specific settings with CFD and a risk assessment at a larger scale in a variety of scenarios.

\section{Methods \label{sec:methods}}
Assessing the risks of viral spread via respiratory droplets in a crowd requires connecting the macroscopic configuration of the crowd and the activity of the attendants to the microscopic dynamics of droplet propagation. Here, we take up the method of our recent work 
\cite{garcia2021model} to derive the number of new infections caused by an index patient on the basis of field data about crowds and mesoscale models of viral transmission (briefly recalled in Sec.~\ref{sec:meth:fieldTOrisk}), but here we aim to fully bridge the scales by anchoring the mesoscale models in a \emph{bona fide} coarse-graining of microscopic simulations of droplet transport that take into account of ambient air flows (see Sec. \ref{sub:meth:CFD_param} and~\ref{sub:meth:CFD}), instead of resorting to mostly \emph{ad hoc} models. The full algorithm is summarized in Appendix~A. 

\subsection{Assess transmission risks in a crowd: general principle}
\label{sec:meth:fieldTOrisk}
We will assess risks in a variety of crowd scenarios, each corresponding to a video recording collected and analyzed by Garcia \textit{et al.} \cite{garcia2021model}. 
For each scenario (streets, stations, markets and more static scenes such as queues and street caf\'es), 
groups of pedestrians were tracked and, in the non-static scenarios, the infection risks within groups were discarded, assuming that a contagious individual is more likely to have infected the people walking in their company elsewhere.

For a given scene, one of the pedestrians, denoted by index $i$ is supposed to be contagious and expel virus-laden droplets. The algorithm is run once for each pedestrian of the scene to gather statistics.  Under the independent action hypothesis  \cite{druett1952bacterial,zwart2009experimental}, each inhaled virus has the same probability to cause an infection, independently of the others. The transmission risks, expressed as a number $C_{i}^{(\tau_{i})}$ of new cases that agent $i$ transmitted to the pedestrians $j$ that crossed his/her path in the interval $[t_{0},t_{0}+\tau_{i}]$ during which he/she was filmed (his/her group $ G_{i}$  excluded, except at the caf\'es; there were no groups in the other static scenario, the waiting line at a screening center), can then be calculated using a Wells-Riley-like equation \cite{SzeTo2010review}:
\begin{equation}
C_{i}^{(\tau_{i})}=\sum_{j\notin G_i}S_{j}^{0}\cdot\left(1-e^{-N_{ij}}\right),\label{eq:C_tau_i}.
\end{equation}
Here, 
 $N_{ij}=\int_{t_{0}}^{t_{0}+\tau_{i}+\tau_{max}}\nu_{ij}(t)\,dt$
is the cumulative transmission risk \cite{tupper2020event}, with $\nu_{ij}(t)$ the instantaneous rate of transmission between the infected person $i$ and a susceptible person $j$. Manifestly, $\nu_{ij}(t)$ is the key quantity of the model and it will be the focus of the next paragraphs.  $S_{j}^{0}$ is the probability
that $j$ is susceptible (i.e., \emph{not already} infected) at the beginning of the observation interval. Assuming $S_{j}^{0}=1$ gives an upper bound on $C_{i}^{(\tau_{i})}$; prior interactions with $i$ in the scenario may reduce $S_{j}^{0}$ below 1, but $C_{i}^{(\tau_{i})}$ cannot become smaller than a lower bound, in practice very close to the upper bound \cite{garcia2021model}. Finally, note that droplets emitted in the interval $[t_{0},t_{0}+\tau_{i}]$ may take some time to reach a susceptible individual and be inhaled after the end of the interval; in practice, a maximum delay $\tau_{max}=21\,\mathrm{s}$ is imposed between emission and inhalation. 

In order to compare different scenarios, $C_{i}^{(\tau_{i})}$ is recast into a rate of infections per hour: $C_{i}\hat{=} C_{i}^{(\Delta T)}=\frac{\Delta T}{\tau_{i}}C_{i}^{(\tau_{i})}$
with $\Delta T=1\,\mathrm{h}$, with the assumption that the recorded videos are representative. Note that in static scenarios (the caf\'es and waiting lines), the treatment is slightly different to account for the fact that interactions occur with a limited number of people, and always the same during the recording. Eq.~\ref{eq:C_tau_i} is applied with $S_{j}^{0}=1$ (everyone except the infected person are susceptible) and the hourly rate is directly computed by using $N_{ij}=\frac{\Delta T}{\tau_{i}}\int_{t_{0}}^{t_{0}+\tau_{i}}\nu_{ij}(t)\,dt$.

\subsection{Instantaneous rate of transmission between two individuals}

Let us consider the droplets emitted by an infected person $i$ (the emitter $E$) and inhaled by another person $j$ (the receiver $R$). At time $t$, $R$ may inhale droplets emitted at different times. In the model, a double decomposition in time is performed. The emission time interval coincides with agent $i$'s observation period  $[t_{0},t_{0}+\tau_{i}]$, and for each emission at time $t_e$, droplets may be inhaled or `received' at time $t_r \in [t_{e},t_{e}+\tau_{max}]$. 

\begin{figure}[ht]
    \begin{center}
	\includegraphics[width=0.5\textwidth]{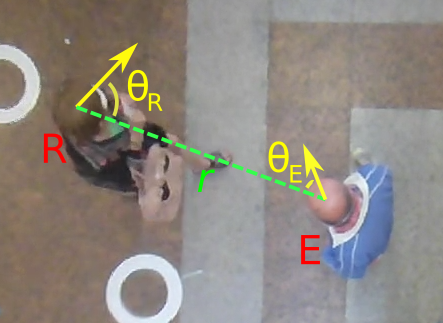}
    \caption{Snapshot of a single interaction filmed at a train station, with an alleged emitter $E$ and a receiver $R$. $r$ is the separation distance and $\theta_{E}$ and $\theta_{R}$ are the directions of emission and inhalation with respect to the line of connection. }
    \label{fig:geometry_transl}
    \end{center}
\end{figure}

More precisely, the instantaneous transmission rate due to droplets emitted at $t_e$ and inhaled at $t_r>t_e$ is expressed as
\begin{equation}
\nu(t_e,t_r)=T_0^{-1}\,\tilde{\nu}\Big[r,\theta_{E}(t_e),\theta_{R}(t_r),\,t_r-t_e,\,\mathrm{ambient\ flows},\,\mathrm{activity(t_e)}\Big]
\label{eq:nu_def}
\end{equation}
where the characteristic time for infection $T_0 \propto n_{\mathrm{inf}}/c_v$ is related to the specifics of the disease (namely, the viral titer $c_v$ in the respiratory fluid and the minimal infectious dose $n_{\mathrm{inf}}$), whereas the function $\tilde{\nu}$ accounts for the fluid dynamics of droplet emission and transport. $r$ is the horizontal distance between the individuals' heads and  $\theta_{E}$
and $\theta_{R}$ are the orientations of the emitter's and receiver's heads, respectively,
relative to the direction of the vector that connects them (see Fig.~\ref{fig:geometry_transl}). While simple \emph{ad hoc} expressions for the function $\tilde{\nu}$ were proposed in \cite{garcia2021model}, here we strive to compute $\nu(t_e,t_r)$ thanks to resolved CFD. 

To do so, we suppose that on each frame in which $E$ is visible he/she emits a small set of droplets, whose evolution is then tracked for $t \in [t_e;t_e+\tau_{max}]$, or equivalently over time delay $\tau \in [0;\tau_{max}]$. For $\tau \leq \tau_{max}$, the spatio-temporal field of viral concentration that $R$ may inhale (depending on his/her position and head orientation)  needs to be known; it will be expressed in the lab frame centered on $E$'s position at the instant of emission $t_e$, as a function of $\tau$, $r$ and $\theta_{E}$.  Of course, the concentration map depends on several parameters: some of them can be extracted from the field measurements, such as the emitter's walking velocity vector $\vctor{v_m}$ and the orientation of the emission (or equivalently the angle between the head and the walking direction); others are unknown, namely, the wind velocity $\vctor{v_w}$ (direction and magnitude) and the characteristics of the exhalation, and will be left as free parameters, whose effect will be assessed.


\subsection{CFD database: Parametrization}
\label{sub:meth:CFD_param}
Were our computational means truly unbounded, we would run one computation for each set of the parameters and each micro-environment, and construct a concentration map for each of these. This is not computationally feasible in the real world.
To circumvent this problem, we build a finite database of CFD simulations and concentration maps in a limited number of situations, from which the required maps will be derived through a suitable change of variables (under certain approximations) or interpolated.

\begin{figure*}[ht]
    \begin{center}
	\includegraphics[width=1.0\textwidth]{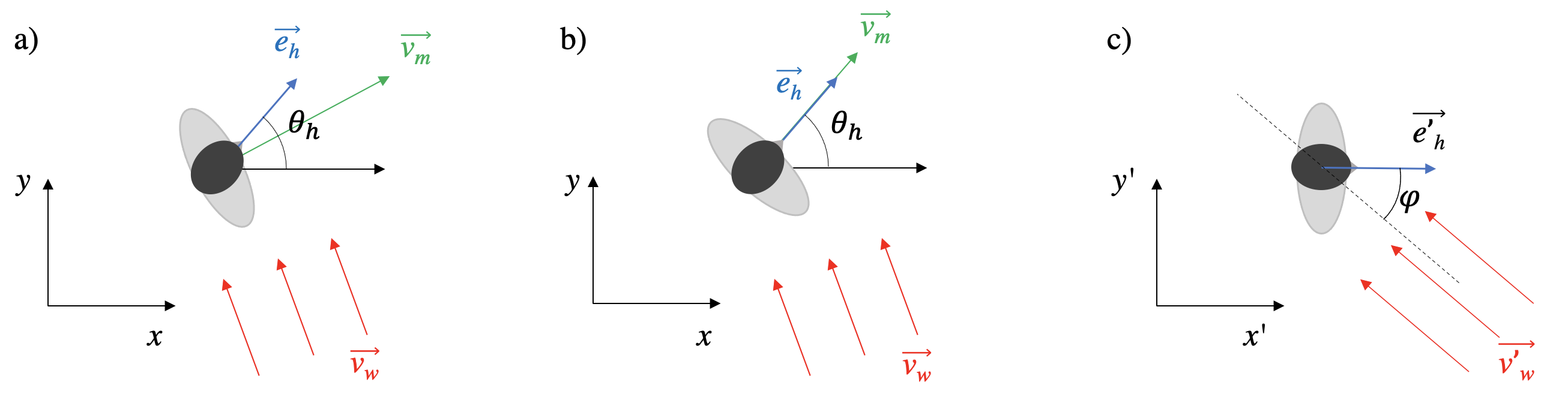}
    \caption{Definition of orientations and velocities relative to the emitter, in the frame of the image (a,b) and in the co-moving frame (c). Panel (b) illustrates the idealized case in which the the walking direction is made to align with the head orientation.}
    \label{fig:frames}
    \end{center}
\end{figure*}

Let ($x$,$y$) be the earthbound frame centered on the emitter $E$ at time $t_e$ and let $\vctor v_{m}$ be $E$'s vectorial velocity (walking) at that time $t_e$ and $\vctor v_{w}$ be the wind velocity; we denote $\vctor e_m$ and $\vctor e_w$ the corresponding unit vectors (directions). Note that the head direction  $\vctor e_h$ does not necessarily align with $\vctor e_m$ in practice, for instance during a conversation, as illustrated in Fig.~\ref{fig:frames}(a). 
We make the following assumptions:
\begin{enumerate}
\item The wind is uniform in space and constant in time during the relevant delay after the emission; velocity gradients in the height direction are neglected. As the relevant transport occurs at the height of human heads, neglecting the boundary layer profile is not expected to impact the results.
\item A walker's motion is a plain translation in the walking direction, at constant speed; idiosyncrasies of the human gait are neglected. 
\item The walking direction is aligned with the head orientation, so that $\vctor v_{m}= v_{m} \vctor e_h$, as represented in  Fig.~\ref{fig:frames}(b). 

Our empirical data show that this assumption is inaccurate, but the associated angular differences are fairly small, with a standard deviation of 26$^\circ$, and, above all, dwindle with the walking speed (see Fig.~\ref{fig:stdTheta}), so much so that they reach  the experimental uncertainty (19$^\circ$) \cite{garcia2021model} for $v_m \geq 1\,\mps$. Large deviations are mostly observed for static people, for which $v_m \approx 0$ and the assumption is theoretically justified. (We have checked that relaxing this assumption of favor of the opposite one, namely, aligning the emissions with the walking direction, leads to similar results, \emph{except} in the static scenarios; see Fig.~\ref{fig:SI_histo_WalkingDir}). 
\end{enumerate}

Under such assumptions, the wind ($\vctor{v_w}$) and the walk ($\vctor{v_m}$) play a symmetric role: as far as air flows are concerned, walking at $1\,\mps$ is equivalent to facing a head wind of $1\,\mps$.
Accordingly, CFD simulations are performed in the frame $(x',y')$ attached to the emitter, located at $(x',y')=(0,0)$, with the basis vector $\vctor{e_{x'}}$ coinciding with the head orientation $\vctor{e_h'}$. This choice entails that the co-moving frame is rotated by an angle $\theta_h=(\vctor{e_x},\vctor{e_h})$  and translated at a constant speed $\vctor v_m$, with respect to the lab frame.
In the co-moving frame, sketched in Fig.~\ref{fig:frames}(c), the wind blows with a velocity $\vctor{v'_w}=(\vctor{v_w}-\vctor{v_m})$; it is modeled by imposing a uniform velocity field $\vctor v_w'$, parallel to the ground, as a boundary condition in the far field. Note that $\vctor v_w'$ may result from the emitter's motion, the wind, or from both of them. 
Finally, we denote $\varphi$ the angle between $-\vctor v_w'$ and the direction of emission $\vctor e_{x'}$. Thus, $v=||\vctor v_w'||$ and $\varphi$ fully parameterize the CFD database.
For instance, in this database, the above situations in which an emitter walks at $1\,\mps$ in still air and  a static emitter faces a headwind blowing at $1\,\mps$ both  correspond to $\varphi=0$ and  $v=1\,\mps$.

\subsection{CFD database: simulation details}
\label{sub:meth:CFD}
The CFD simulation protocol is detailed in Appendix~\ref{app:num}. In short, a still-standing manikin mimics a man who is breathing through the mouth, at a rate of 20 breaths of $1\,\mathrm{L}$ per minute, i.e., one breath every 3 seconds, with an equal time for exhalation and inhalation. This signal was originally designed to replicate the breathing flow rate of a walking person, but also applies for speaking. Large-eddy simulations are performed along the same lines as \cite{Abkarian2020speech,garcia2021model}, using the incompressible version of the Navier-Stokes equations, which was found to provide the best compromise between cost and accuracy of the simulations. Each simulation starts with 3~cycles to establish the flow, followed by 4 cycles during which statistics are collected.

Echoing Abkarian \textit{et al.} \cite{Abkarian2020speech}, we remark that the unsteady starting jets close to the mouth tend to form a main jet whose characteristics far from the mouth resemble those of a steady jet, with a limited influence of the details of the exhalation signal. This is why we use the same aerodynamic simulations to model mouth breathing and speaking.
However, the number and sizes of emitted droplets will differ between breathing and speaking.

Overall, the database consists of 25 microscopic simulations for ambient relative velocity $v\,(\mps) \in \mathcal{S}_{v} \hat{=} \{0,\,0.1\,0.3,\, 1.0,\, 2.0\,\}$ and $\varphi\in \mathcal{S}_{\varphi} \hat{=} \{0,\frac{\pi}{6}, \frac{\pi}{3}, \frac{\pi}{2}, \frac{3\pi}{4}, \pi\}$, 
plus the case $v=0.0\,\mps$, for which $\varphi$ is undetermined (the simulations at $v=0.1\,\mps$ were used for control exclusively). The simulation cost increases with $v$. For $v=2.0\,\mps$, a simulation takes more than $\approx 80$ hours on 5 AMD EPYC Rome 7H12 bi-sockets nodes (640 cores) of the IRENE-AMD partition of Joliot-Curie cluster (TGCC/CEA, France).

`Test-particles', i.e., droplets of diameters uniformly distributed in $d_p\in[0.1\,\mathrm{\mu m}, 1 \,\mathrm{mm}]$, are injected into the airflow exhaled by the  emitter. Importantly, the number of injected droplets (about 60,000 per breath) is not intended to be consistent with the empirical data for breathing~\cite{asadi2020efficacy,alsved2020p}, but merely to collect sufficient statistics in terms of particles behavior over a few cycles; since these droplets have very weak mutual aerodynamic interactions in the puff, at any reasonable concentration, this statistical contrivance will play virtually no role in the results.

Indeed, any \emph{actual} distribution of emitted droplet sizes can be obtained by suitably resampling the simulated distribution, that is, assigning appropriate weights to the droplets depending on their size so as to match the desired distribution.
In practice, following \cite{johnson2011modality}, superpositions of log-normal distributions of droplet diameters $d_p$, whose cumulative functions $P_s$ obey
\begin{equation}
dP_s(d_p) \propto \frac{1}{\ln(\sigma)}\cdot  e^{-\frac{1}{2} \Big(\frac{\mathrm{ln}(d_p/\bar{D})}{\ln(\sigma)}\Big)^2 }\,d\mathrm{ln}d_p.
\end{equation}
Breathing features only one such mode, with $\bar{D}=0.8\,\mathrm{\mu m}$ and $\sigma=1.3$, whereas vocalization (speech) features one mode at $\bar{D}=0.8\,\mathrm{\mu m}$ and $\sigma=1.3$, multiplied by a coefficient 0.069, and one mode at $\bar{D}=1.2\,\mathrm{\mu m}$ and $\sigma=1.66$, multiplied by a coefficient 0.085. The third mode associated with speech is peaked at $\bar{D}=217\,\mathrm{\mu m}$, with $\sigma=1.795$ and a coefficient 0.001; it thus corresponds to large droplets unlikely to be inhaled (see Sec.~\ref{sub:inhalation}). The coefficient associated with the breathing mode is found by recalling that breathing produces approximately 20 times fewer droplets than normal speech \cite{asadi2020efficacy}. 

Note that the foregoing sizes correspond to those measured by Johnson et al. \cite{johnson2011modality} prior to the application of their corrective factors, which notably account for evaporation. Indeed, small respiratory droplets are expected to undergo quick partial evaporation, which makes it sensible to consider their propagation with their evaporated diameters;
besides the (slight) effect on droplet transport, we expect no further incidence of the application of a constant corrective factor on the diameters, thanks to our renormalization with the characteristic infection time.

\subsection{Coarse-grained dynamic maps of viral concentration \label{sub:methods_cg_maps}}

Once resampled, the detailed configuration of the simulated droplets is coarse-grained into
dynamic maps of viral concentration $c(r,\theta,\tau)$, where $(r,\theta)$ are polar coordinates in the earthbound frame centered on the emitter's mouth \emph{at the instant of emission} $t_e$ and $\tau$ is the delay since emission of the droplets. $\theta=0$ is the direction of emission. 
These maps $c(r,\theta,\tau)$ are 
obtained by binning droplets in space and time, with a resolution $\delta r=20\,\mathrm{cm}$ on $r$, $\delta \theta=\frac{\pi}{12}$ on $\theta$, and $\delta \tau=0.2\,\mathrm{s}$ on the delays $\tau$, i.e., the lifetimes of droplets, and then
computing the total volume of droplets in each spatiotemporal cell, within a $40\,\mathrm{cm}$-thick horizontal slice centered on the emitter's mouth (the yellow box displayed in Fig.~\ref{fig:micro_puffs}), and dividing it by the cell volume. This relies on the
 assumption that viral copies are homogeneously distributed in respiratory fluids and each raises the same risk of infection (regardless of the droplet size), which is classical in modeling but possibly underestimates the viral load in small particles \cite{Wang2021airborne,fennelly2020particle}. 
 The resulting maps are then symmetrized with respect to the $\theta=0$ axis ($\vctor{e_h}$), if  such a symmetry is expected to hold, i.e., for head and tail winds. Finally, for any relative wind velocity $\vctor{v}$, the emitter can be assigned a walking speed $v_m$ without additional CFD simulations, by simply translating the origin of the concentration maps with the time delay $\tau$ at a speed $v_m$ in the direction opposite to the emitter's head orientation, i.e., along $-\vctor{e_h}$. 
 Note that the question of the normalization of the concentration maps, i.e., their scale, will be circumvented by setting a characteristic infection time $T_0=15\,\mathrm{min}$ for someone standing face-to-face, a distance $r_c=50\,\mathrm{cm}$ away from a static \emph{speaking} emitter. 
 In other words, the quantum is defined as the quantity of virus inhaled in 15 minutes while standing 50~cm away from a static speaking person, without wind; this volume of droplets is used to normalize the transmission risks.

\section{Results: Transmission risks generated by an emitter \label{sec:results_micro}}

In this section, we inspect the spatiotemporal pattern of risks (i.e., virus-laden aerosols) emitted by an index patient, depending on the environmental conditions.

\subsection{Propagation of the droplets simulated with CFD}
\begin{figure*}[ht]
    \begin{center}
	\includegraphics[width=1.0\textwidth]{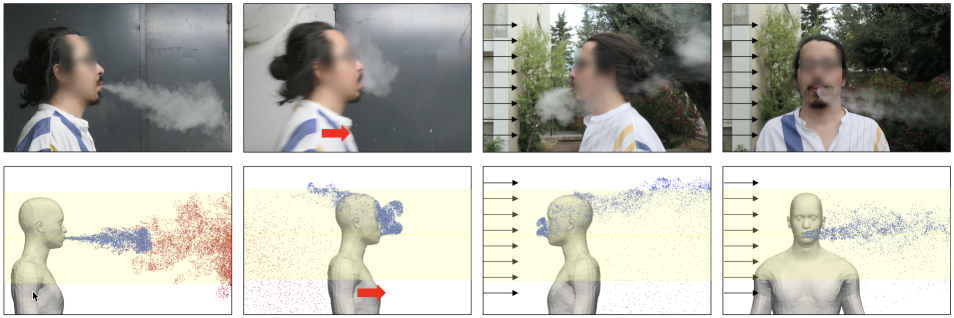}
    \caption{Examples of the variation of the exhaled puffs with the wind and the walking speed. (Top row) Photographs. (Bottom row) CFD simulations, displayed at time 0.75s of the 6th cycle, when the exhaled flow is close to its maximum.  Droplets of less than 10 microns in diameter are displayed to mimic tracer particles. They are colored in blue for those exhaled over the 6th cycle (0.75 s before and less) and in red for those exhaled during cycles 1 to 5. The region with a yellow overlay materializes the box within which droplets will be counted (see the main text). \\
    First column: exhalation in still air. Second column: exhalation during walk (at $1.0\,\mathrm{m\cdot s}^{-1}$ in the CFD) in still ambient air. Third and fourth columns: exhalation in headwind and crosswind, respectively (at $2.0\,\mps$ in the CFD). Pictures are illustrative, as walking speed and wind speed do not match, and exhalation has not been characterized.}
    \label{fig:micro_puffs}
    \end{center}
\end{figure*}

Figure~\ref{fig:micro_puffs} shows the propagation of an arbitrary number of droplets of less than 10 microns in diameter, emitted by the manikin in the CFD
simulations, for distinct incident velocities ($v=0.0$, $1.0$ and $2.0\,\mps$) and angles $\varphi$. Interestingly, both the relative wind (generated by walking) and the external wind have a dominant effect on the propagation, after a short first regime during which the direction of emission prevails. This holds true even at very low wind speeds, which would not even be qualified as `light breeze' (6-11 km/h) on the Beaufort scale. For example,  a wind of $1.0\,\mps$ (3.6 km/h) corresponds to the air flow felt when walking in a still environment.
To illustrate the consistency of these numerical results, one subject gave his consent to be photographed while smoking an e-cigarette. This experiment consisted in several exhalations indoors while walking or not, and outdoors in the wind, whose speed was not measured. 
Thus, the walking speed and the wind speed could  \emph{not} be matched between the experiments and the simulations; the comparison mostly has an illustrative purpose. 
The whole experiment lasted a few minutes and did not lead a substantial change in the subject's consumption of his e-cigarette.

Although qualitative, the comparison highlights the singularity of the case without relative wind (left), where a long jet may develop without being perturbed. Another observation is the similarity of the pictures/results obtained while walking in still air (2nd column) or being static in headwind (3nd column): the spatial extent of the puff in front of the subject is substantially smaller than in the static case without wind (4th column) and the puff disperses behind the subject's head.  
We observe that the wind may transport droplets farther over shorter periods of time, but we will need to turn to concentration maps to concentration maps to determine if this heightens transmission risks.

\subsection{Dynamic maps of viral concentration}

Following the coarse-graining method exposed in Sec.~\ref{sub:methods_cg_maps}, the detailed output of the CFD simulations is converted into dynamic maps of viral concentration, that is, spatiotemporal diagrams of risks centered on the emitter
at the moment when droplets are shed. 

\begin{figure*}[ht]
    \begin{center}
	\includegraphics[width=1\textwidth]{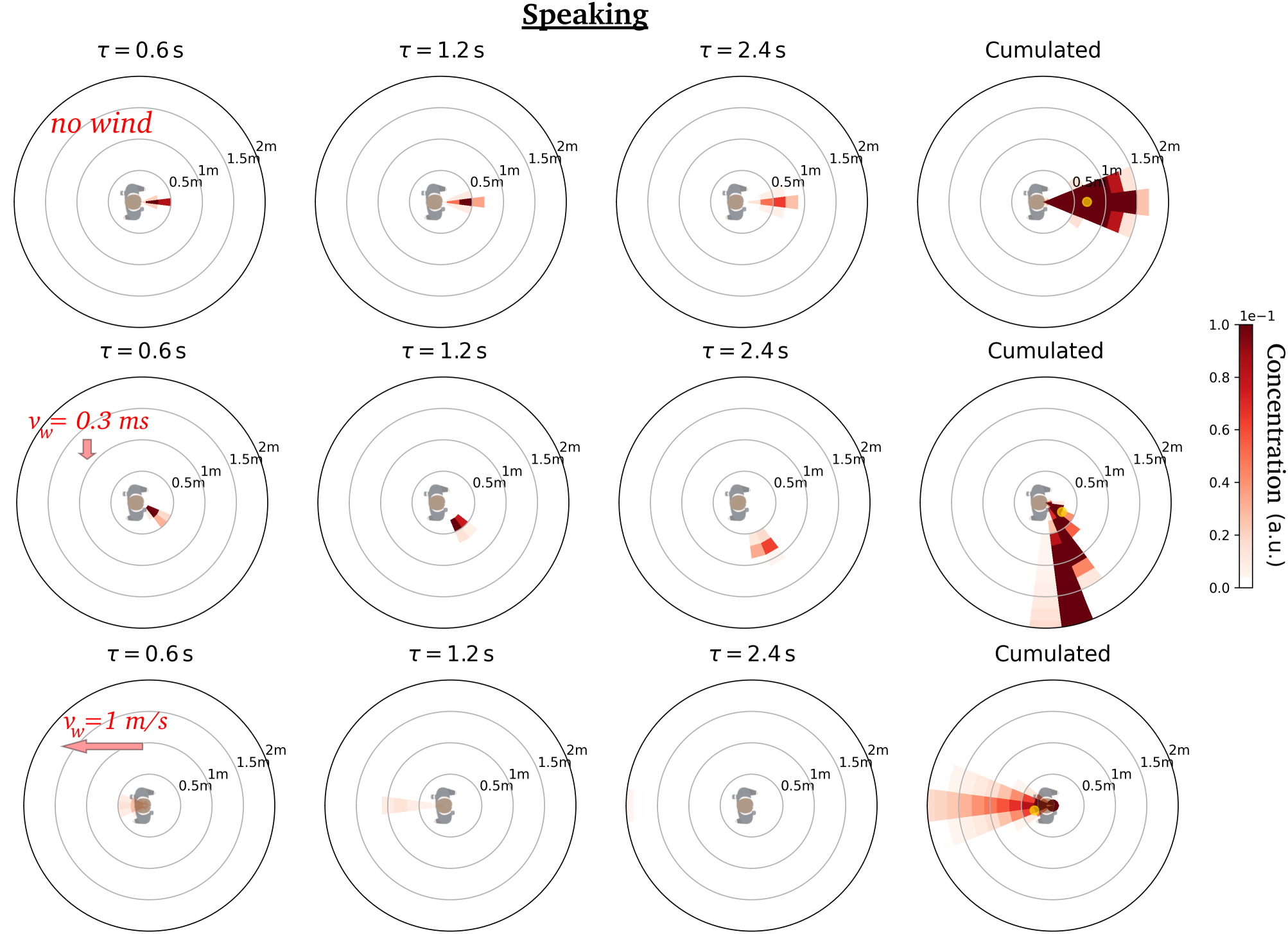}
    \caption{Dynamic maps of viral concentration associated with talking : Effect of the external wind. (Top row) no wind, (middle row) lateral wind blowing at $0.3\,\mps$, (bottom row) head wind blowing at $1\,\mps$. The emission is arbitrarily aligned along the $x$-axis and the color bar saturates at the arbitrarily imposed top value.}
    \label{fig:diagrams_wind}
    \end{center}
\end{figure*}

We first comment on the maps obtained at zero walking speed $v_m$, shown in Fig.~\ref{fig:diagrams_wind}. The values displayed in these diagrams, say at position $(r,\theta)$ and after a delay $\tau$, for a relative wind $\vctor{v'_w}$, can naturally be interpreted as the risks incurred by a non-moving receiver located at $(r,\theta)$ relative to static emitter under an external wind $\vctor{v'_w}$, up to an inhalation coefficient, but let us mention that it also describes the situation in which the emitter and the receiver are walking at the same velocity $\vctor{v_m}$ when the wind blows at $\vctor{v_w}=\vctor{v'_w}+\vctor{v_m}$.
We notice, once again, that in the first fractions of a second the puff tends to follow the line of emission, but it is then steered
by the wind, while remaining fairly compact, and it is swept more than 2~m away from the emitter in a matter of seconds at any finite wind speed ($v\geqslant 0.3\,\mps$), compared to more than a dozen seconds in windless conditions. This further underlines the singularity of the windless case as far as one is concerned with droplet transport. 

Not surprisingly, maximal risks are incurred in the immediate vicinity of the emitter in the time-cumulative diagrams (where the peak for $r>20\,\mathrm{cm}$ away from the emitter is shown as a small yellow dot), but the azimuthal position is influenced by the relative wind. For example, it lies behind the emitter for a head wind blowing at $1.0\,\mps$ [Fig.~\ref{fig:diagrams_wind}(bottom)].

\begin{figure}[ht]
    \begin{center}
	\includegraphics[width=0.9\columnwidth]{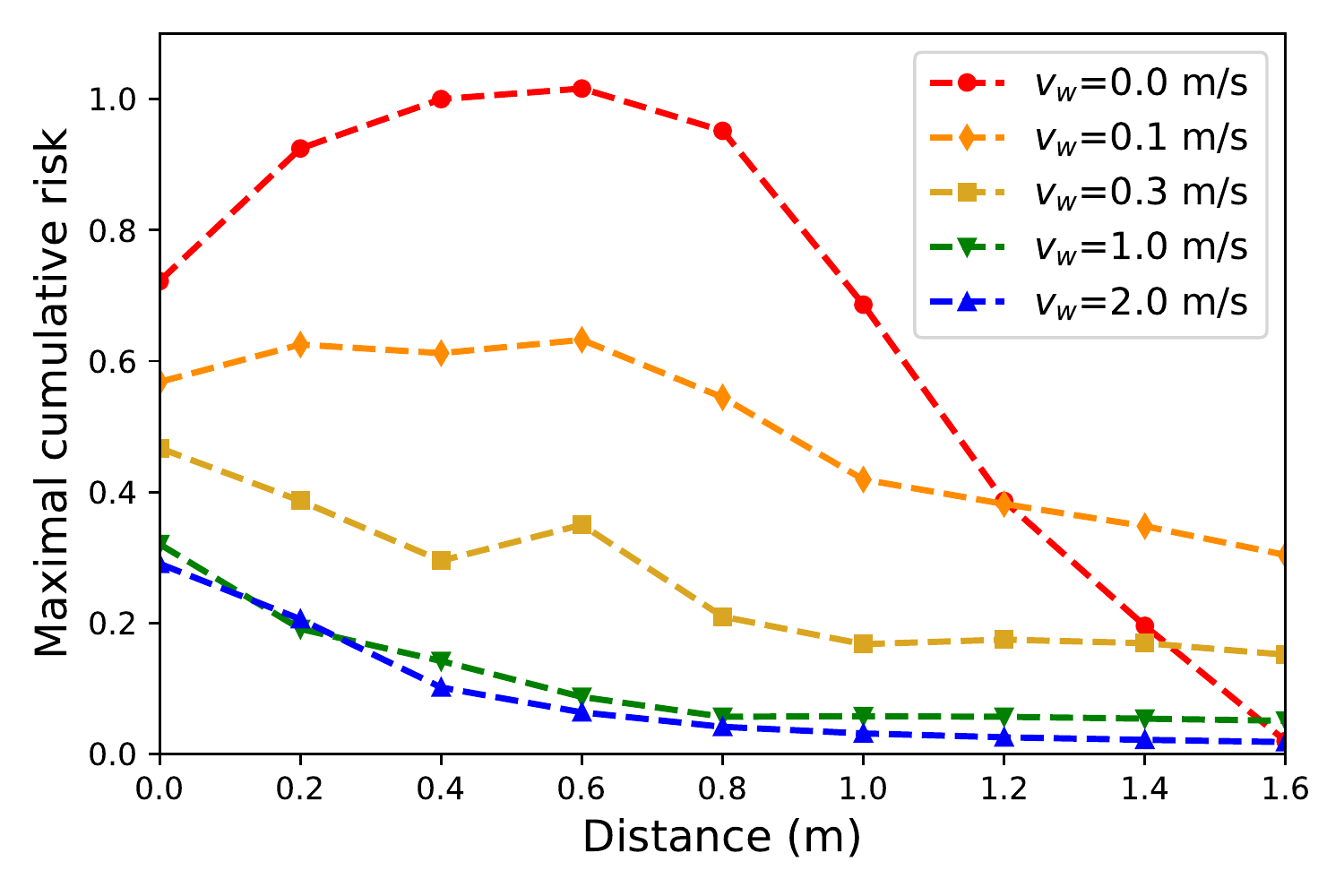}
    \caption{Radial decay of the maximum concentration of viral particles emitted by a static pedestrian over all azimuthal directions, in the cumulative concentration maps associated with speaking. An external wind is assumed to blow perpendicularly to the head orientation ($\varphi=\frac{\pi}{2}$), at the speed $v_w$ indicated in the legend. The concentrations were normalized to one at a distance $r_c=50\,\mathrm{cm}$ in windless conditions. The color scale is saturated to make the low levels of concentration visible.}
    \label{fig:radial_decay}
    \end{center}
\end{figure}

Besides, it is now apparent that the major effect of the wind, besides steering the puff, is to quickly disperse the emitted droplets and thus to lower transmission risks.
We illustrate this in Fig.~\ref{fig:radial_decay} in the case of a side wind
by plotting the radial decay of the maximum concentration of viral particles over all azimuthal directions for different wind speeds. A wind speed as low as $0.3\,\mps$ reduces the peak value at 50-cm distance by a factor of 3, roughly speaking. Even the slightest, almost imperceptible draft, at $v_w=0.1\,\mps$, has a significant effect on the transmission risks. This demonstrates how singular are the stagnant air conditions ($v_w=0$) often used to model droplet transport. 
The local maximum observed in the $v_w=0$ case is notably due to inhalation, which removes the last droplets exhaled from the near-mouth region. Such droplets are swept out in the presence of wind.

\begin{figure*}[ht]
    \begin{center}
	\includegraphics[width=1\textwidth]{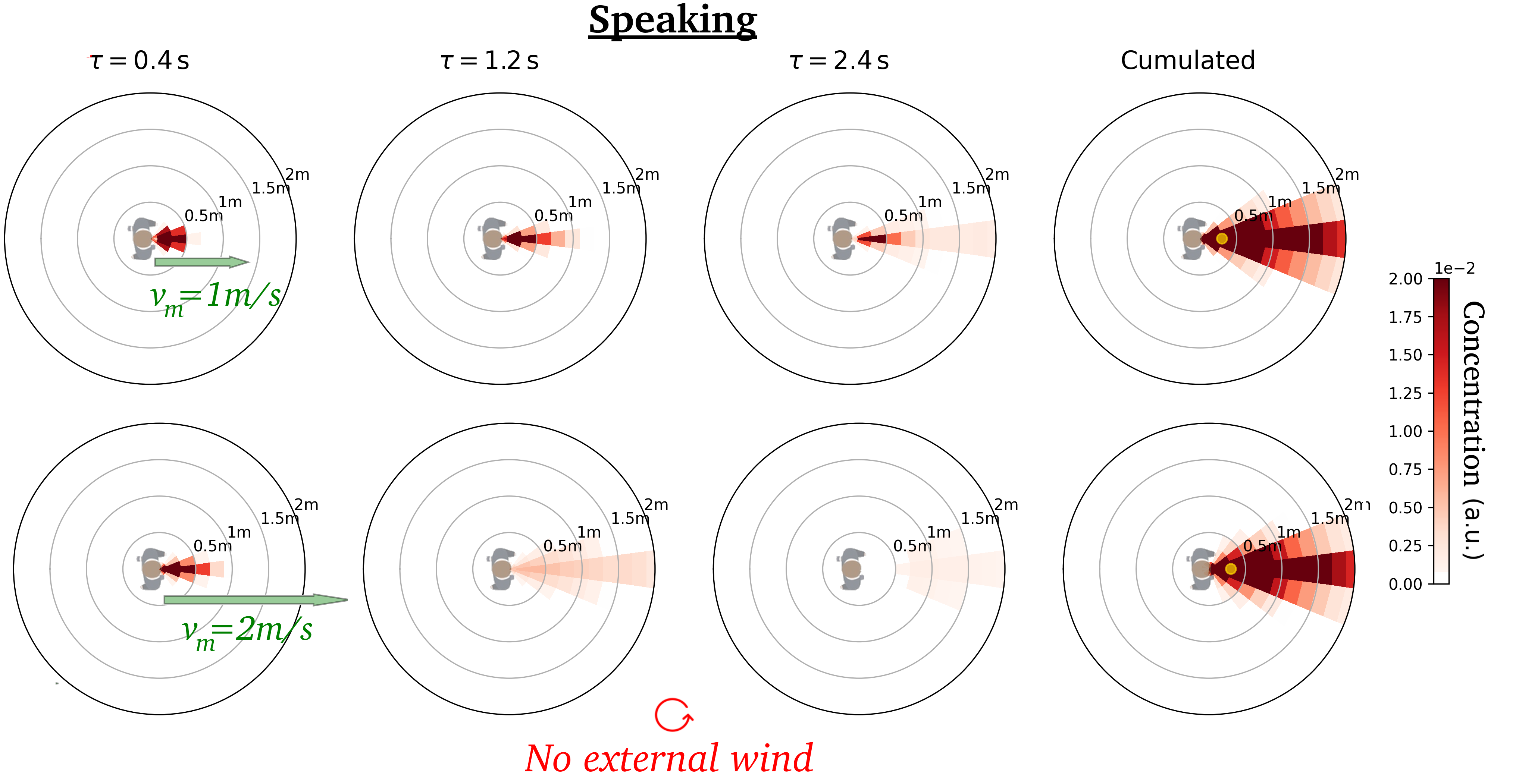}
    \caption{Dynamic maps of viral concentration associated with speaking (in the lab frame): Effect of the walking speed in windless conditions. Walking speed $v_m=$ (top) $1.0\,\mps$, (bottom) $2.0\,\mps$}
    \label{fig:diagrams_walk}
    \end{center}
\end{figure*}

Let us now make the emitter move while breathing or speaking. Figure~\ref{fig:diagrams_walk} shows that the  air flow generated by walking drags the puff forward, along the emitter's path; similarly to an external wind, this drag also tends to sweep away the emitted droplets. Incidentally, recall that the concentration maps are shown in the Earth's frame, and not in the walker's co-moving frame, which can explain why no clear detachment transition of the puff is observed as the walking speed increases \cite{li2020effects}.

\subsection{Robustness to simulation details related to the flow dynamics}
In view of the many recent reports highlighting the sensitivity of droplet transport to physical and numerical details, before proceeding with the assessment of risks, we examine the influence of these details on our results (more information in Appendix \ref{app:algo}, notably in Fig.~\ref{fig:SI_compDiagrams_CFD}):

(i) Moderate variations of droplet sizes have little impact on droplet propagation. Indeed, the concentration maps obtained for subsets of droplet sizes between $0.5\,\mathrm{\mu m}$ and $20.0\,\mathrm{\mu m}$ are qualitatively very similar; this is due to the fact that results are gathered on regions of 40 cm height, while sedimentation velocity of $20.0\,\mathrm{\mu m}$ droplets is of the order of the centimeter per second \cite{Netz2020physics}. In addition, the Stokes number remains small for such droplets, which means that particles essentially follow the air flow.

(ii) Buoyancy effects, due to the different temperature of the puff compared to the ambient air, have an impact, but a moderate one, as we see when these effects are introduced into CFD simulations. The buoyancy generated by the thermal plume surrounding human bodies (which are hotter than the environment) \cite{sun2021human} has not been directly simulated. As stressed by Nielsen and Xu in a recent review \cite{Nielsen2022multiple}, the air inhaled indoors mainly comes from the lower part of the body and is lifted to the nose by the thermal plume. To quantify its effect, we have compared the reference concentration maps with their counterparts measured when the region of interest is shifted downwards by 20~cm, with limited differences.

(iii) Evaporation does not affect the results much. Crucial in this lack of incidence is the fact that, because of their finite fraction $c$ of non-water content, droplets do not fully evaporate, but shrink into residues (also called droplet nuclei), of final size of the order of $c^{\frac{1}{3}}$ times their initial size \cite{Netz2020physics,Seyfert2022stability}.

All these effects are already weak with no external wind, but are even further reduced in the presence of wind, which always ends up dominating transport far from the mouth.

\subsection{Interpolation of spatiotemporal diagrams}
\label{sub:interpolation}
Given that the CFD database only contains a limited number of cases $(v,\varphi)\in \mathcal{S}_{v} \times \mathcal{S}_{\varphi}$, interpolation is needed to obtain the spatiotemporal diagram corresponding to given relative wind conditions $(v,\varphi)$ at a delay $\tau$. We select the two closest values $\varphi_1$ and $\varphi_2$ ($\varphi_1\leqslant \varphi \leqslant \varphi_2$) in $\mathcal{S}_{\varphi}$ and $v_1$ and $v_2$ ($v_1\leqslant v < v_2$) in $\mathcal{S}_{v}$. Now, since $\varphi$ controls the rotation of the exhaled air puff under the wind and $v$ affects the propagation dynamics, a na\"ive linear interpolation over $\varphi$ and $v$ would perform poorly.

Instead, noticing 
from Fig.~\ref{fig:diagrams_wind} and its kin that the puff is first transported in the direction of emission, and then in the wind direction, we handle (very) short delays $\tau \leqslant \tau^{\star}(v) \hat{=} 0.12/v$ distinctly from longer delays $\tau > \tau^{\star}(v)$. For the former, the directions of emission are already aligned with $\vctor{e'_x}$ in the spatiotemporal diagrams, so no rotation is needed. For the latter, the diagrams corresponding to $\varphi_i$, $i=1,2$, will be rotated by an angle $\varphi - \varphi_i$ prior to interpolation. 

Turning to the speed variable $v$, noticing that (for a given wind direction) increasing $v$ has an effect somewhat comparable to `fast-forwarding the movie', i.e., inspecting the diagram at a \emph{shorter} delay $\tau$, the diagrams corresponding to speeds $v_i$, $i=1,2$,  are probed at delays $\tau_i=\tau \, v/v_i$ (if $v_i>0$; $\tau$ otherwise) and their values (which represent the transmission risk over a fixed time interval $\delta \tau$) are rescaled by multiplication with $v_i/v$.

\begin{figure*}[ht]
    \begin{center}
	\includegraphics[width=1.0\textwidth]{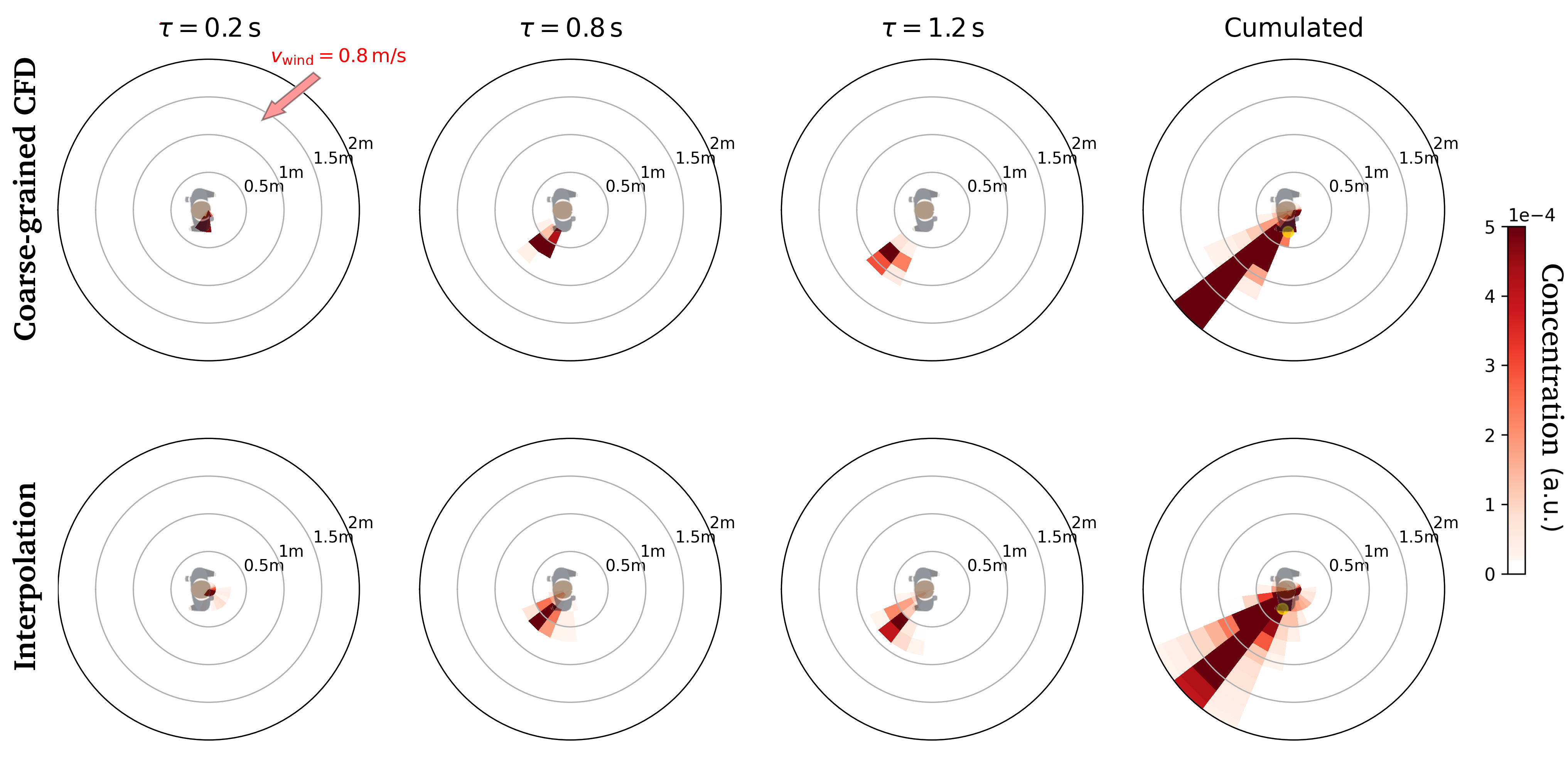}
    \caption{Interpolation of dynamic concentration maps at for $v=0.8\,\mps$ and $\varphi=\frac{\pi}{4}$. (Top) Coarse-grained map calculated using a \emph{bona fide} CFD simulation corresponding to these specific conditions. (Bottom) Concentration map obtained by interpolation.}
    \label{fig:diagrams_interp}
    \end{center}
\end{figure*}

To sum it up, for each $v_i$, we interpolate linearly between the diagrams corresponding to $\varphi_1$ and $\varphi_2$, after suitably rotating them if $\tau > \tau^{\star}$, and linear interpolation between the resulting diagrams at $v_1$ and $v_2$ yields the final diagram. No interpolation is needed on the walking speed $v_m$ because, from the CFD output at 
$(v,\varphi)$, we are able to generate and store dynamic maps for a wide range of walking speeds (in practice, $v_m=0,\,0.1,\,0.2,\, \dots,\,2\,\mps$).
The example shown in Fig.~\ref{fig:diagrams_interp} (and Fig.~\ref{fig:SI_diagrams_interp}), in which a genuine concentration map computed for  $\varphi=\frac{\pi}{4}$ is compared to its interpolated counterpart, demonstrates that this method produces quite reasonable results.

\subsection{Inhalation and the case of large oral droplets}
\label{sub:inhalation}
Once respiratory droplets have been emitted, they must be inhaled by a receiver to bring on a risk of transmission. 
So far the dynamic concentration maps have been established irrespective of the receiver and her head orientation. 
We now take care of the latter by inserting a multiplicative factor $\nu_R$ accounting for inhalation in the transmission rate function $\tilde{\nu}$ in Eq.~\ref{eq:nu_def}, on top of the concentration at the receiver's location \cite{Nielsen2022multiple}. We consider two possibilities;
in the first one the incident puff can be inhaled only if it `hits' the side of the receiver's head containing the mouth and nose, hence
\begin{equation}
\nu_R(\theta_R) = 
    \begin{cases}
 1\text{ if }\theta_R\in [-\frac{\pi}{2},\frac{\pi}{2}],
\\ 0\text{ otherwise.}
\end{cases}
\label{eq:nuR_aniso}
\end{equation}
where $\theta_R$ was defined in Fig.~\ref{fig:geometry_transl}.
The second option, isotropic inhalation, is inspired by the steady-flow situation in which the concentration becomes homogeneous all around the head, including the `dead-waters' zone located downstream, so that
\begin{equation}
    \nu_R(\theta_R) = 1.
\label{eq:nuR_iso}
\end{equation}

In doing so, we assume that the breathing activity of the receiver has little incidence on the transport of the droplets produced by the emitter. While it is true that under specific face-to-face conditions the receiver's expiration can significantly perturb the emitter's expiration flow (and possibly act as a shield) \cite{Giri2022colliding}, more generally, this assumption sounds reasonable, especially during the receiver's inhalation, provided that the emitter is not too close to the receiver.

Besides, the droplets must be small enough for inhalation to be possible. This will always be the case for the breathing and vocalizing modes in the micron range, but the question is worth discussing for the third mode (`oral' mode), peaked at $217\,\mathrm{\mu m}$. Indeed, the typical sedimentation speed for a droplet of diameter $d_p=100\,\mathrm{\mu m}$ (resp. $d_p=217\,\mathrm{\mu m}$) is $v_g\approx 0.3\,\mps$ (resp. $1.5\,\mps$), which is larger than the magnitude of inhalation velocity measured by Murakami \cite{Murakami2004analysis}, for instance. This is confirmed by our own simulations of nose breathing presented in Appendix~\ref{sec:SI_large_droplets} (see Fig.~\ref{fig:isospeed} in particular). 
Therefore, inhalation of the droplets of that mode is deemed rather unlikely and in the following we will focus on the smaller droplets. Nevertheless, in Appendix~\ref{sec:SI_large_droplets} the possibility to inhale these larger droplets is restored and leads to a distinct global picture.

\section{Results: Risks of new infections at the macroscale \label{sec:results_macro}}

Moving on to macroscopic crowds, we now couple the coarse-grained dynamic maps of viral concentration obtained in the previous section with field data about crowds in daily-life situations.

\subsection{Empirical crowd dynamics}
\begin{figure}[ht]
    \begin{center}
	\includegraphics[width=0.8\columnwidth]{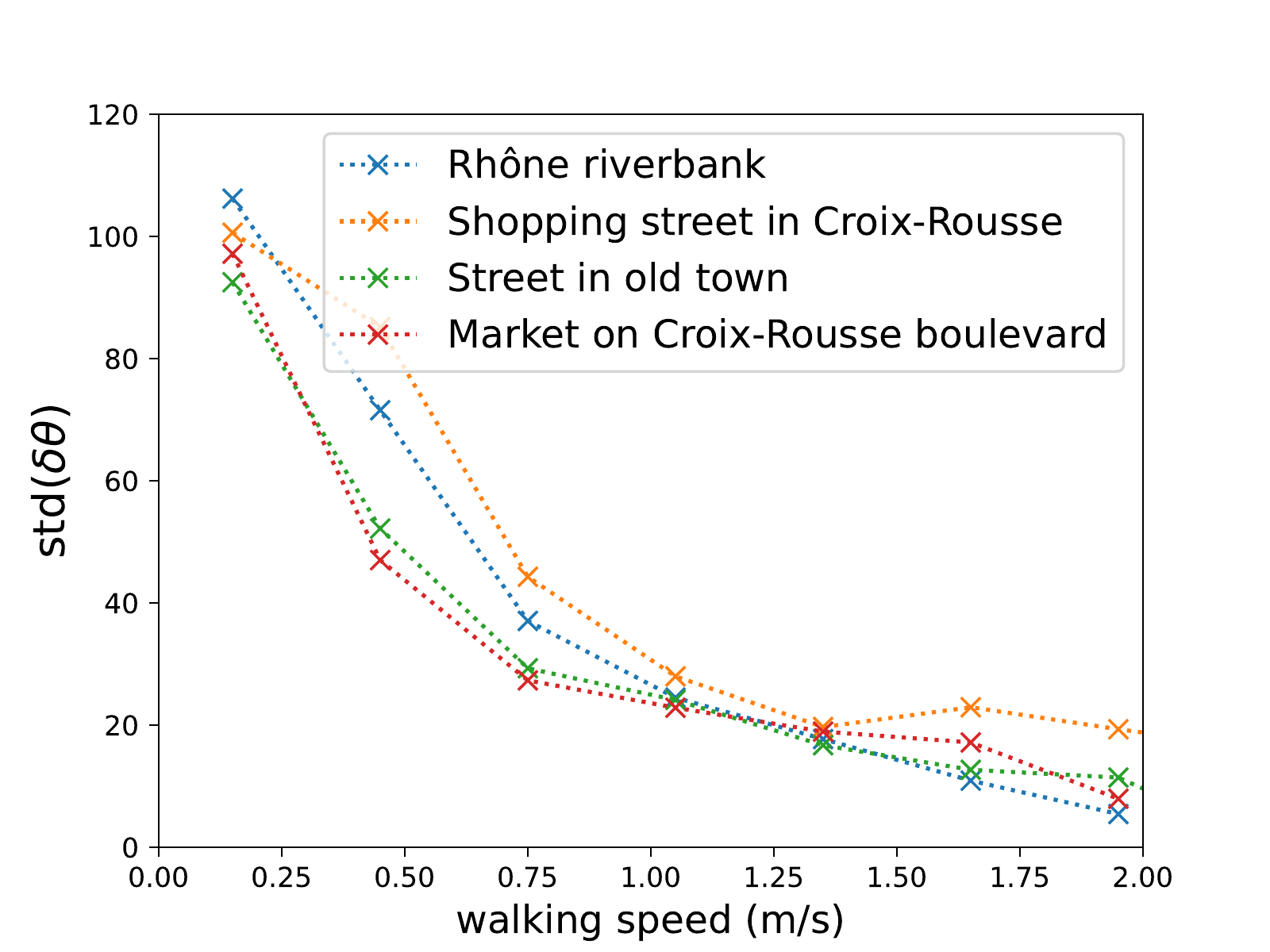}
    \caption{Standard deviation of the difference between the head orientation $\vctor{e_h}$ and the walking direction $\vctor{e_m}$, as a function of walking speed.}
    \label{fig:stdTheta}
    \end{center}
\end{figure}

We begin with a presentation of the empirical scenarios that will be used as test-cases in our macroscopic risk assessments. They are listed in Tab.~\ref{tab:scenarios} and include fairly busy streets (and riverbanks), metro and train stations, an outdoor market, and street caf\'es. All scenarios are in outdoor settings or in large, well ventilated areas and the data were collected in the metropolitan area of Lyon, France, during the COVID-19 pandemic, between July 2020 and January 2021. Note that these data are openly available on the Zenodo platform \footnote{https://zenodo.org/record/4527462}. More details about the data acquisition and pedestrian tracking protocols can be found in Garcia et al. \cite{garcia2021model}. In short, the pedestrians' positions and head orientations are marked on the videos every 0.5~s; the temporal resolution is then increased to a point every 0.1~s through linear interpolation. By double tracking pedestrians, the experimental uncertainty on the absolute positions was estimated to below or around 20~cm while the error on the head orientations had a standard deviation of $19^{\circ}$. 
The accuracy of the angular data is well evidenced by our ability to capture the tendency of pedestrians to look more and more straight ahead as they walk faster, as reflected in Fig.~\ref{fig:stdTheta} by the variation of the 
standard deviation of $\delta \theta$, the angular difference between the head orientation and the walking direction, with the walking speed.

Finally, to compensate for the narrow field of view and the interactions thus missed, a reweighting process, based on an estimation of the number of missed contacts, was proven to effectively correct the bias towards shorter-ranged interactions \cite{garcia2021model}.

\begin{table*}[h]
\noindent 
\begin{center}
\begin{tabular}{| c | c | c |}
\hline 
Scenario / Location  & \#ped & \textbf{mean density (m$^{-2}$)}\tabularnewline
\hline 
\hline 
Pedestrian banks of the RhÃŽne river, close to the Morand Bridge & 164 & \textbf{0.042}\tabularnewline
\hline 
\textcolor{black}{Plaza in front of Perrache hub
(Hall - Level 1)} & 1021 & \textbf{0.038}\tabularnewline
\hline 
Part-Dieu train station - Ground level/passage area (indoors) &  875 & \textbf{0.22}\tabularnewline
\hline 
Busy street - Under the Passerelle Bouchut &  800 & \textbf{0.05}\tabularnewline
\hline 
Bellecour subway station - Platform of Line D (indoors) & 849 & \textbf{0.26}\tabularnewline
\hline 
Croix-Rousse boulevard - street caf\'es & 13 & \textbf{/}\tabularnewline
\hline 
Grande rue de la Croix Rousse (shopping street) &  420 & \textbf{0.06}
\tabularnewline
\hline 
Saint-Jean street in the Old Town of Lyon & 481 & \textbf{0.11}\tabularnewline
\hline 
Place des Terreaux - Bar/Restaurant terraces & 30 & \textbf{/}\tabularnewline
\hline 
Croix-Rousse - Main market alley & 183 & \textbf{0.46}\tabularnewline
\hline 
COVID-19 testing site & 66 & \textbf{/}\tabularnewline
\hline 
\end{tabular}
\par\end{center}
\caption{\label{tab:scenarios}Scenarios under study; \#ped denotes the number of tracked pedestrians. All sites are in the metropolitan area of Lyon, France; most are outdoors.}
\end{table*}

\subsection{Different perspectives for the assessment of risks}

We now apply the methodology exposed in Sec.~\ref{sec:methods} to the field data, recalling that
the static scenarios (queue and street caf\'es) are handled slightly differently from their moving counterparts: People are assumed to keep interacting with the same neighbors over the whole period $\Delta T=1\,\mathrm{h}$ in the former, whereas in the latter they will interact with new people. Moreover, in the moving scenarios, the risks of infecting one's co-walkers are overlooked, because we are interested in the supplemental risks generated by the scenario under consideration and the co-walkers probably interacted with the index patient in more risky places, such as enclosed settings; all the other people are considered susceptible \footnote{The reasoning can straightforwardly be extended to situations in which a fraction of the population is immunized against the virus \cite{garcia2021model}.}. By contrast, no social groups are taken into account at street caf\'es.

Besides, it should be underlined that risks are here quantified by the number of new cases $\bar{C}^{(\Delta T)}$ expected in each setting when an index patient is present on the premises for a duration $\Delta T=1\,\mathrm{h}$, and \emph{not} the total rate of new infections in the scenario or the risks incurred by a typical attendant. We claim that this perspective, centered on the infected person, is
the relevant one at the \emph{collective} scale, for policy-making: It enables the decider to compare the infection potentials of the different activities in which an infected person would engage. In this sense, and contrary to the total rate approach, a massive gathering will be deemed to present higher risks than a number of smaller gathering only if it leads to higher $\bar{C}^{(\Delta T)}$ than in the smaller gatherings.
Compared to the perspective centered on the receiver, it is true that, if a given fraction $\gamma$ of the crowd is contagious (irrespective of the scenario), 
the infecter-centered risk assessment also reflects the \emph{average} risk per hour
incurred by an \emph{individual} in the crowd, $\gamma \bar{C}^{(\Delta T)}$, but the two perspectives may have very dissimilar distributions (for instance with a low risk for a large number of people or a high risk for a small number of people). We put the focus on the infecter-centered risk assessment.

\subsection{Rate of new infections in perfectly windless conditions}

\begin{figure*}[ht]
    \begin{center}
	\includegraphics[width=1.0\textwidth]{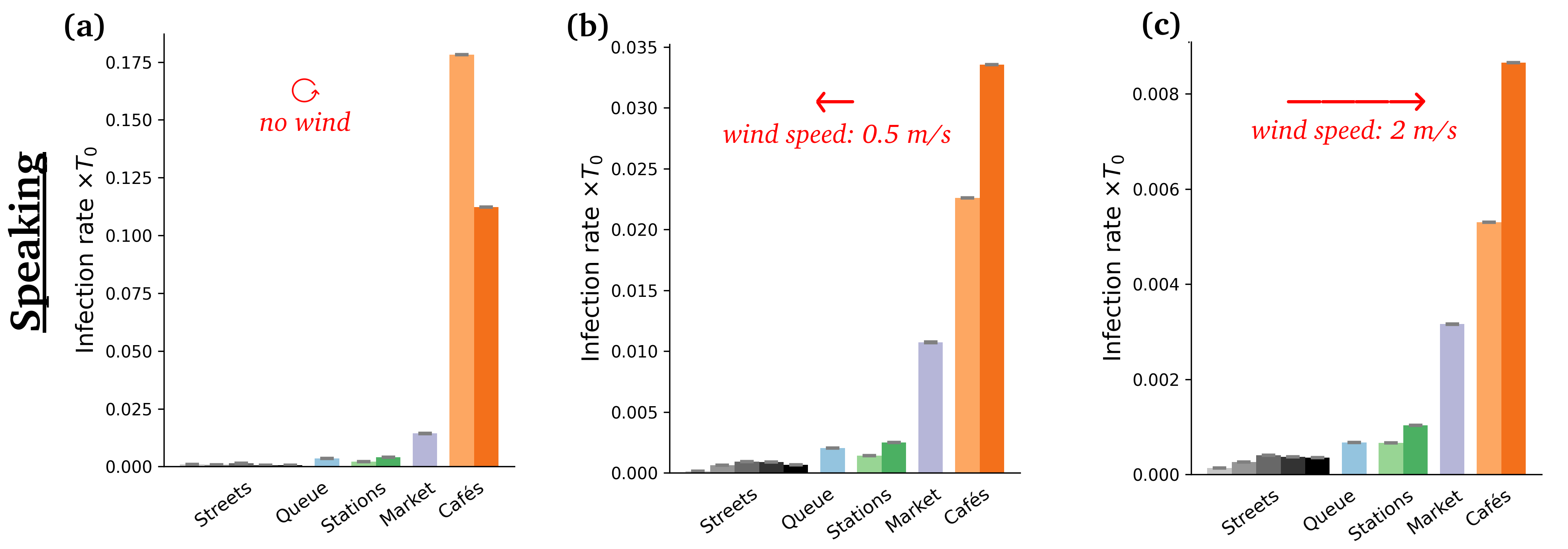}
    \caption{Estimated risks of infection associated with speaking in each scenario, for different wind speeds. Pay attention to the widely different scales in each panel.}
    \label{fig:histo1}
    \end{center}
\end{figure*}

Figure~\ref{fig:histo1}(a) shows the rates of new infections in windless conditions for the different scenarios, when all pedestrians are supposed to be constantly talking.  caf\'es present the highest risks by far, followed by the outdoor market and, further down, the metro and train platforms and halls (filmed in the midst of the pandemic), whereas the risks raised by fairly busy streets are comparatively quite low.
Reassuringly, these trends are perfectly in line with those found previously, on the basis of various \emph{ad hoc} models, which totally discarded the relative winds generated by walking, among other aspects.  But, here, the account is more quantitative, given that our transmission models are rooted in fluid dynamical models; the only adjustable variable is the characteristic time of infection $T_0$ (once rescaled by this time, the rates of infections vary little with $T_0$, within reasonable bounds).

Switching to breathing through the mouth instead of talking does not affect the ranking in the slightest way, which makes sense, given the relative insensitivity of droplet transport to small variations in droplets sizes (Fig.~\ref{fig:SI_compDiagrams_CFD}). But it dramatically lowers the risks, by a factor of order 100, consistently with the lower volume of respiratory droplets produced in this case.

In reality, people will carry out a mix of respiratory activities; the risks should then be computed by an average of the risks raised for each type of activity, weighted by the proportion of time spent for each activity; breathing through the nose may be considered to raise no risks.
In practice, this weighting will further enhance the risks associated with caf\'es, insofar as talking will probably occupy a larger fraction of time in this scenario than in the other ones.

\subsection{Effect of modest winds or ambient air flows}
Introducing an external wind alters the foregoing picture to some extent (Fig.~\ref{fig:histo1}b-d). Most strikingly, the absolute levels of risks are strongly depressed, e.g. by a factor of 4 in the case of the market, for a wind speed of $v_w=1\,\mps$. This is in line with the dispersal effect of wind established in Fig.~\ref{fig:radial_decay}.
Besides, the risk gap between street caf\'es and the outdoor market lessens with increasing $v_w$, so much so that the two scenarios become about as risky under external winds blowing at $2\,\mps$, which is still calm air. (Recall however that this comparison only holds if similar expiratory activities are performed in all scenarios; otherwise, the activities should be reweighted). This effect is easily understood as the wind bends the particularly unfavorable propagation of droplets in the case of face-to-face conversations, and favors transmission in isotropically dense settings. The distribution of rates of new infectious caused by the different individuals that were filmed is presented in Fig.~\ref{fig:boxplot}.

How robust are these results to variations in the simulated conditions? In principle, at equal wind speed, the wind direction is not expected to impact the results. This is verified in most scenarios, but not all. Some effect is observed for one of the street caf\'es and the queue at the screening center: In these cases, the specific settings that we filmed displayed
preferential directions (whether it is the alignment of the table or that of the queue), which may couple with the wind direction.

Turning to the directional dependence of inhalation, the anisotropic inhalation coefficient $\nu_R$ given by Eq.~\ref{eq:nuR_aniso} was used so far. 
Replacing it with a fully isotropic one, Eq.~\ref{eq:nuR_iso}, tends to enhance the risks, since it allows more directions for inhalation. This is most acute in the windless queuing scenario, where the face-to-back alignment of people in the queue used to make inhalation inefficient.

\begin{figure*}[ht]
    \begin{center}
	\includegraphics[width=1.0\textwidth]{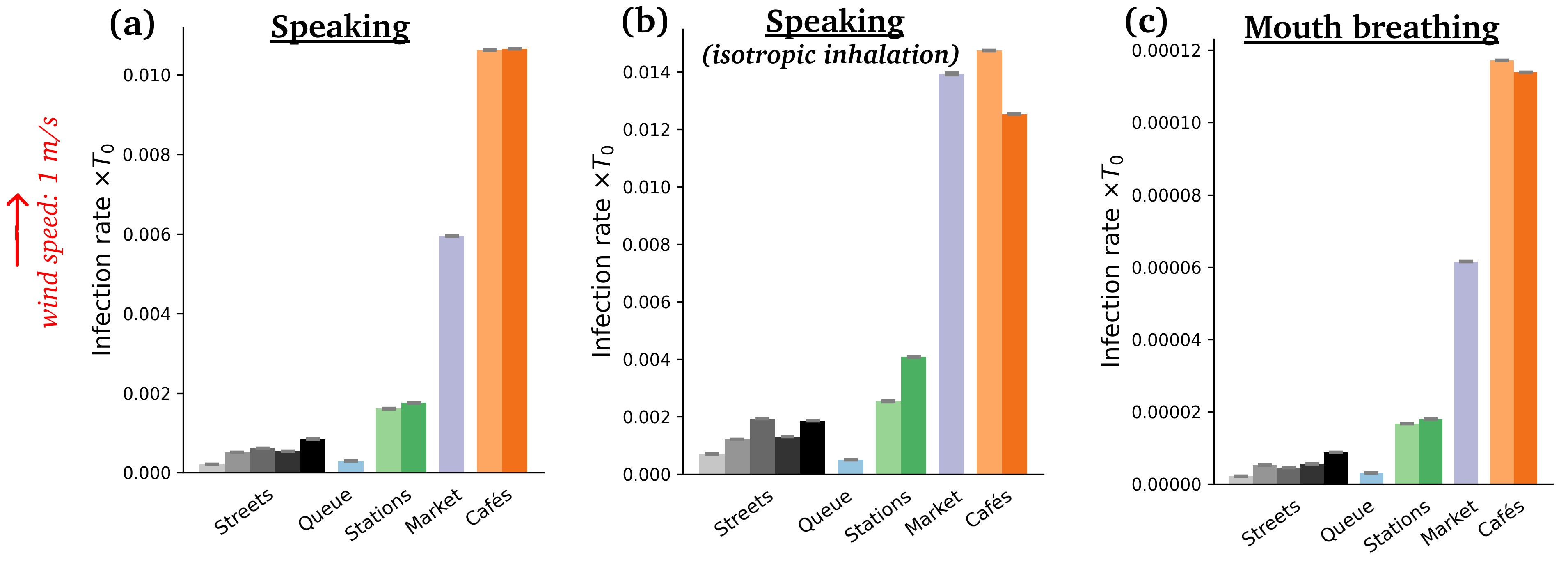}
    \caption{Estimated mean risks of infection in each scenario under an external wind blowing at $1\,\mps$, for (a) speaking, (b) speaking, with an isotropic inhalation coefficient, (c) breathing through the mouth. Pay attention to the widely different scales in the panels.}
    \label{fig:histo2}
    \end{center}
\end{figure*}

\begin{figure}[ht]
    \begin{center}
	\includegraphics[width=0.9\columnwidth]{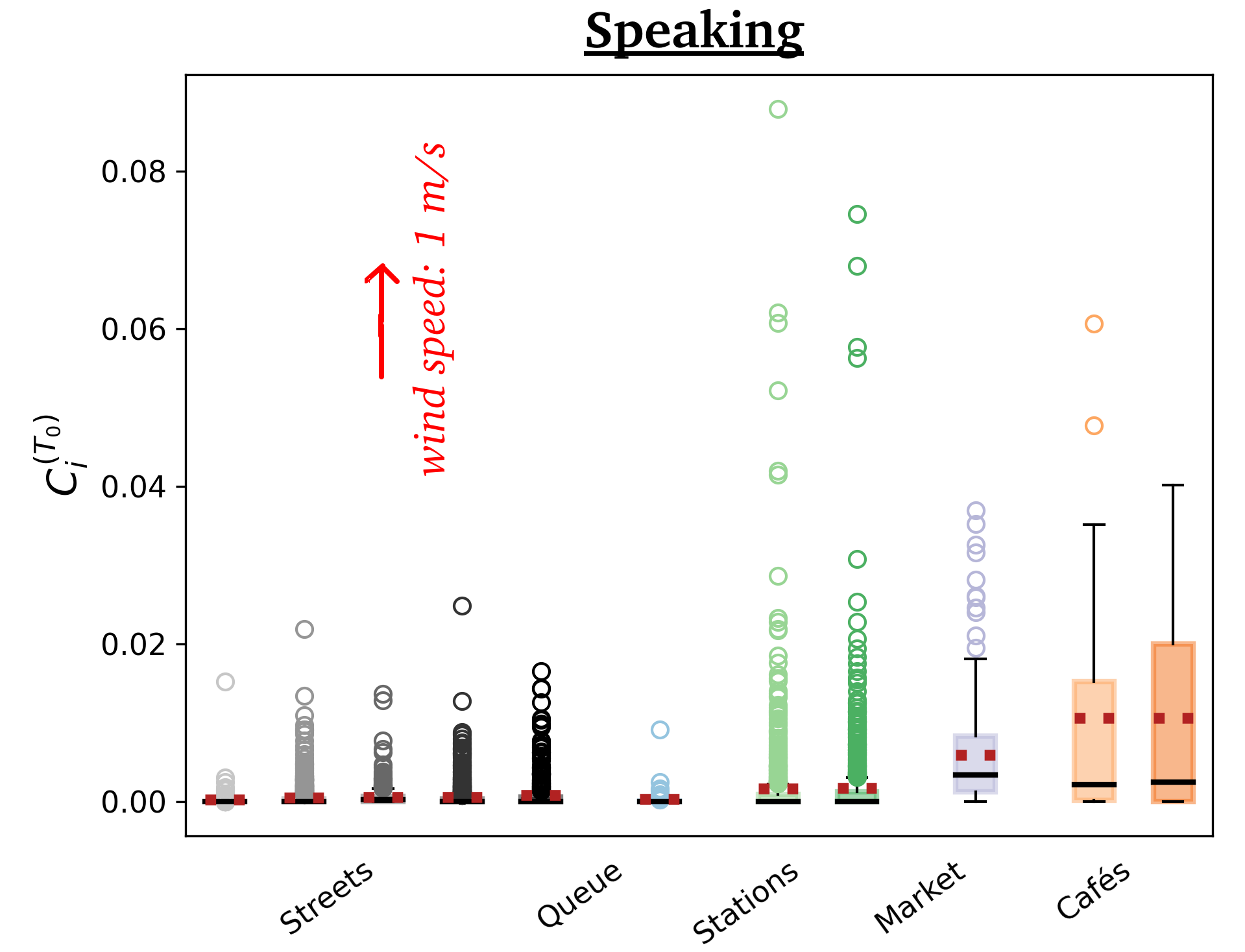}
    \caption{Box plot of risks of new infections associated with speaking under an external wind of $1.0\,\mps$ blowing to the 'North'. The dashed red lines represent mean values, solid back lines are medians and open symbols are outliers.}
    \label{fig:boxplot}
    \end{center}
\end{figure}

\subsection{A practical guide to assessing risks in real or test scenarios}
We close this section with a succinct summary of the practical way in which risks of viral transmission can be assessed using our framework:
\begin{itemize}
    \item Collect or simulate the $(x,y)$-trajectories of people in the scenario; the head orientations can be measured or assumed to align with the walking direction;
    \item If needed, interpolate between points in order to increase the temporal resolution;
    \item Feed these trajectories into the Python scripts coupling them to the dynamic concentration maps to get the mean rates of new infections $\bar{C}^{(\Delta T)}$ for each expiratory activity and wind velocity under study
    \item Estimate the fraction of time $\alpha_t$ spent talking and the fraction of time $\alpha_b$ spent breathing through the mouth;
    \item Compute the rate of new infections as a weighted average
    \footnote{Note that, in the moving scenarios, the average should be performed on the rates of new infections $\bar{C}^{(\Delta T)}$ whereas, in the static scenarios, they should be performed on the number of inhaled droplets $N_{ij}$ defined in Sec.~\ref{sec:meth:fieldTOrisk}.}
    of the $\bar{C}_{\alpha}^{(\Delta T)}$ over the expiratory activities $\alpha$ (if talking is not negligible, then $\bar{C}^{(\Delta T)}\simeq \bar{C}^{(\alpha_t\,\Delta T)}_{\mathrm{talking}}$) at the typical external wind speed $v_w$, or averaged over the distribution of different wind speeds and wind directions.
\end{itemize}

\section{Conclusions}

In summary, we have put forward and implemented a methodological framework to assess the risks of viral transmission via 
short-ranged exposure in crowds. It provides an unprecedented connection between the fluid dynamical propagation of respiratory droplets at the microscale and field-data about macroscopic crowds, using spatio-temporal maps of viral concentration.

These concentration maps give insight into the transmission risks in binary interactions, highlighting the paramount effect of the respiratory activity, owing to the much larger volume of respiratory droplets expelled while talking, as compared to mouth-breathing; the effect of (even very modest) winds or ambient air flows is also apparent. The impact of air flows had been reported previously in case studies of specific (mostly indoor) settings \cite{li2020evidence,vuorinen2020modelling,Poydenot2021risk,mariam2021cfd,mills2021review,rivas2022impact}; its generality is underscored here. Our study further shows how the walking velocity of pedestrians also contributes to decreasing the risks, through the same mechanism. 

By coupling these binary transmission risks with pedestrian trajectories, the rates of new infections generated
by a contagious individual wandering for one hour on the premises was assessed. We put this to the test using field data collected in diverse daily-life scenarios, which were thus ranked by risks. Consistently with our previous findings with much coarser models, street caf\'es present
the highest risks among the investigated scenarios in windless conditions, due to the configuration of the crowd (even if the
larger fraction of time spent talking is left aside), followed by the observed outdoor market and, further down the list, train and metro stations (at the peak of the pandemic). However, our finer models also stress the dramatic quantitative effect of the wind on these results, which strongly depresses transmission risks and tends to reduce the gap between street caf\'es and the busy outdoor market. The work thus contributes to explaining why overall outdoor settings appear to raise substantially fewer risks of viral transmission than enclosed spaces, besides the negligible risk of long-range air-borne transmission in non-confined settings. 

Note that the critical influence of air currents even at low speed also urges one to reconsider short-range transmission risks indoors. Indeed, indoor drafts have a typical speed of a few tens of centimeters per second \cite{Liu2007neural}, which implies that they can dominate droplet propagation after the exhaled puffs loose their initial momentum. This should be taken into account in microscopic studies of droplet transport, which are usually performed in stagnant air.

To conclude, on the bright side, it is worth noting the generality of the proposed framework, whereby a diversity of scenarios may be tested, e.g., to explore the efficiency of redesign strategies aimed at mitigating viral spread, in a stadium or at any other mass gathering. Besides, it is on no account restricted to COVID-19. On a less positive note,
we must plainly acknowledge that our approach, which provides a unique way to span six orders of magnitude in lengthscales in order to address transmission risks, rests on several serious approximations.
Relaxing them would overcome some limitations of the model, possibly at the expense of more costly simulations.
First, the CFD simulations could be refined to more accurately reflect the exogenous turbulence of the wind (before it interacts with the manikin's body and the expiration flow). Along similar lines, the influence of the receiver's exhalation and inhalation flow on the transport of droplets generated by an emitter \cite{Giri2022colliding} could be taken into account, even though this may require many more CFD simulations.
Height differences between people could also be incorporated, and should be so if the crowd mixes standing and seated people.
To allow other researchers to contribute to these prospective improvements and/or apply the proposed approach, our main scripts have been made publicly available on the GitHub repository \url{https://github.com/an363/InfectiousRisksAcrossScales}; the other scripts can be requested by email.

\section*{Acknowledgments}
The setup of the CFD simulations was designed collectively, with P. B\'enard, G. Lartigue, V. Moureau (CORIA Rouen, France), G. Balarac, P. B\'egou (LEGI Grenoble, France), Y. Dubief (Univ. Vermont, USA) and R. Mercier (Safran Tech, France). We thank our colleague O. Kaplan for generating the smoke jets of Fig.~3. 
CFD simulations were performed using HPC resources
from TGCC-IRENE (Grants No. AP010312425 and A0100312498). AN acknowledges the help of B. Fray during his internship and thanks the MODCOV group for facilitating the initiation of collaborations.
This work was funded by
Agence Nationale de la Recherche: projects SeparationsPietons (ANR-20-COV1-0003, A. Nicolas) and TransporTable (ANR-21-CO15-0002, S. Mendez).

\pagebreak
\appendix 

\setcounter{figure}{0}
\renewcommand\thefigure{S\arabic{figure}}

\section{Algorithm for the calculation of transmission risks}
\label{app:algo}
Algorithm~\ref{alg:risk} presents a pseudo-code for the risk assessment from a series of images. One person is considered as an index patient $i$ in the scenario, then the risk is calculated for each of the other people in the movie potentially interacting with $i$. $i$ is then varied for ensemble averaging. The wind velocity  $\vctor{v_{w}}$  is a free parameter that has not been measured.

\begin{algorithm}
Among the individuals in the movie, select $i$, the assumed infected person

Set the wind velocity $\vctor{v_{w}}$

Initialize $C_{i}^{(\tau_{i})}=0$

\For{p:=1:Number of images}{

    \If{$i$ is in image $p$}{
         Get $t_p$, the time of image $p$
         
         Get $i$'s walking speed $\vctor{v_{m}}$
         
         Get the orientation of $i$'s head  $\vctor{e_h}$
         
         Calculate $v$ and $\varphi'$ and interpolate the concentration maps from the CFD database.
         
        \For{j:=1:People in $i$' environment}{
        
             Initialize the risk for $j$
             
            \For{q:=p+1:Number of images}{
                \If{$j$ is in image $q$}{
                     Get the time of image $q$, $t_q$
                     
                     Calculate the delay $\tau=t_q-t_p$.
                     
                    \If{$\tau \leq \tau_{max}$}{
                         Calculate the relative position of $j$ with respect to the emission $r_j(t_q)-r_i(t_p)$
                         
                         Get the concentration as a function of $r_j(t_q)-r_i(t_p)$ and $\tau$, for the relevant parameter values ($\vctor{v_{w}}$, $\vctor{v_{m}}$, $\vctor{e_h}$).
                         
                         Update the risk for $j$.
                    }
                }
            }
        }
         Update $C_{i}^{(\tau_{i})}$ with $j$'s contribution. 
    }
}
 Calculate $C_{i}^{(\tau_{i})}$. 
 
 Recast $C_{i}^{(\tau_{i})}$ into an hourly rate of transmission $C_{i}$. 
\caption{Algorithm to assess the transmission risk associated with one contagious person in a movie \label{alg:risk}}
\end{algorithm}

\section{Numerical method and simulations details}
\label{app:num}

This appendix details the microscopic numerical simulations used in the paper.

\subsection*{Physical model and flow solver}
Three-dimensional numerical simulations are performed in the idealized case of non-buoyant jets, neglecting temperature effects. To account for turbulence, we use Large Eddy Simulations (LES) as reported and described in \cite{Abkarian2020speech}.
 The fluid flow is governed by the incompressible  versions of the filtered continuity and  Navier-Stokes equations:
\begin{equation}
\frac{\partial \bar{u}_i}{\partial x_i} =0, 
\label{eq:continuity}
\end{equation}

\begin{equation}
\rho\frac{\partial \bar{u}_i}{\partial t} + \rho\frac{\partial \bar{u}_i \bar{u}_j}{\partial x_j} = -\frac{\partial \bar{p}}{\partial x_i}+\mu \frac{\partial^2 \bar{u}_i}{\partial x_j \partial x_j} + \frac{\partial \tau_{ij}}{\partial x_j},
\label{eq:NS}
\end{equation}
where $\bar{u}_i$ is the filtered fluid velocity component in the $i$th direction, $\bar{p}$ the filtered pressure, $t$ the time, $x_i$ the spatial coordinate in the $i$th direction, $\rho$ the constant air density and $\mu$ the constant dynamic viscosity. $\tau_{ij}=\rho(\overline{u_i u_j} - \bar{u}_i \bar{u}_j)$  is the residual stress-tensor coming from the subgrid-scale unresolved contribution, for which a closure needs to be provided. Here we use the so-called sigma model \cite{Nicoud:2011} which has notably been built to yield zero extra dissipation in laminar flows, so that it is well adapted to situations at moderate Reynolds numbers where transition to turbulence occurs~\cite{Nicoud:2018,Zmijanovic:2017}, which is the case here.

The considered fluid is assumed to represent air at constant ambient temperature. The kinematic viscosity is fixed at $\nu=\mu/\rho=1.5\times 10^{-5}$ m$^2$/s.

For the present work, we used the flow solver YALES2 \cite{Abkarian2020speech,Zmijanovic:2017} \footnote{\url{https://www.coria-cfd.fr/index.php/YALES2}}. 
The fluid equations are discretized using a fourth-order finite-volume scheme, adapted to unstructured grids \cite{Moureau:2011a,Moureau:2011b}. 

Exhaled droplets are represented as spherical Lagrangian particles, tracked individually with a point-particle Lagrangian approach. One-way coupling is used as the concentration of droplets is small (of the order of a few particles per liter \cite{asadi2020efficacy}). The droplet motion is obtained by advancing them along the flow
\begin{equation}
\frac{\mathrm{d}\mathbf{x}_p}{\mathrm{d}t} = \mathbf{u}_p,
\end{equation}
where $\mathbf{x}_p$ is the position of the droplet and $\mathbf{u}_p$ its velocity.

Conservation of momentum is given by Newton's second law:
\begin{equation}
\frac{\mathrm{d}}{\mathrm{d}t}\left(m_p \mathbf{u}_p\right) = \mathbf{F}_p^G + \mathbf{F}_p^A,
\end{equation}
where $m_p$ is the mass of the droplet, $\mathbf{F}_p^G$ is the buoyancy force and  $\mathbf{F}_p^D$ the drag force. The buoyancy force and drag force read:
\begin{equation}
\mathbf{F}_p^G = \left( \rho_p - \rho \right)\frac{\pi}{6}d_p^3\mathbf{g} \quad \quad {\rm and} \quad \quad \mathbf{F}_p^D = m_p\frac{1}{\tau_p}\left( \mathbf{u}_p - \mathbf{u} \right),
\end{equation}
where $\rho_p$ is the droplet density, $\rho$ the local gas density, $d_p$ the droplet diameter and $\mathbf{g}$ the gravitational acceleration. 
$\tau_p$ is the characteristic drag time. It is modeled with the empirical correlation of Schiller and Naumann~\cite{SchillerNaumann1935} for moderate values of the droplet Reynolds number. This correlation tends to the Stokes law at low Reynolds numbers. 

\subsection*{Computational domain and grid}

The computational domain is a 3~m-high, 6-m long box in the wind direction; its width is 4~meters. The manikin's head is oriented in the $x'$ direction and the ambient wind blows in a variable direction as a function of the incident angle $\varphi$ (see Fig.~\ref{fig:frames}). 

\begin{figure*}[ht]
    \begin{center}
	\includegraphics[width=\textwidth]{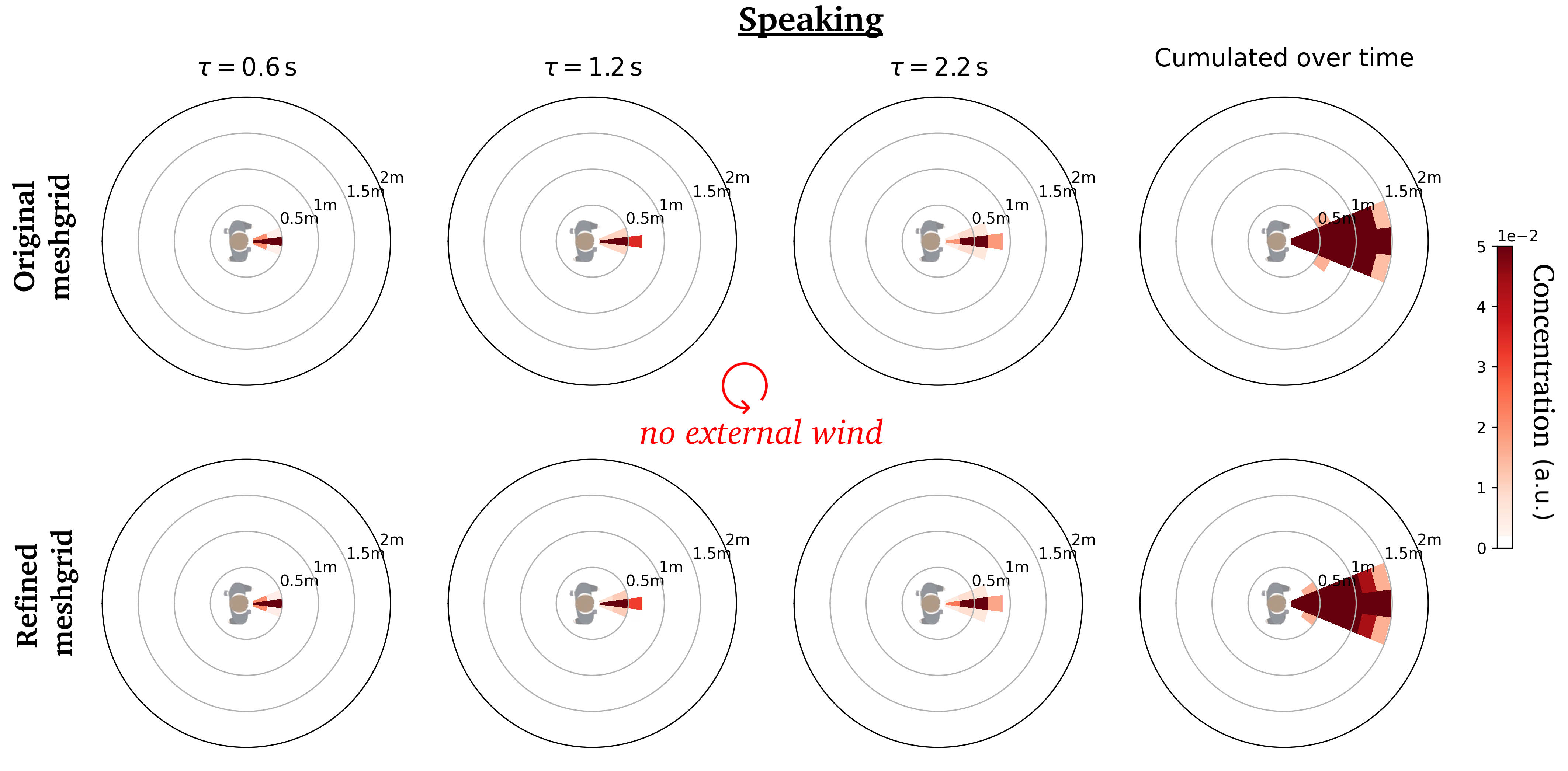}
    \caption{Comparison of the dynamic concentration maps generated  at distinct time delays $\tau$ for a static pedestrian in windless conditions, by coarse-graining CFD simulations performed with (\emph{top row}) the original meshgrid, (\emph{bottom row}) a refined meshgrid.}
    \label{fig:SI_FinerGrid_0ms}
    \end{center}
\end{figure*}

\begin{figure*}[ht]
    \begin{center}
	\includegraphics[width=\textwidth]{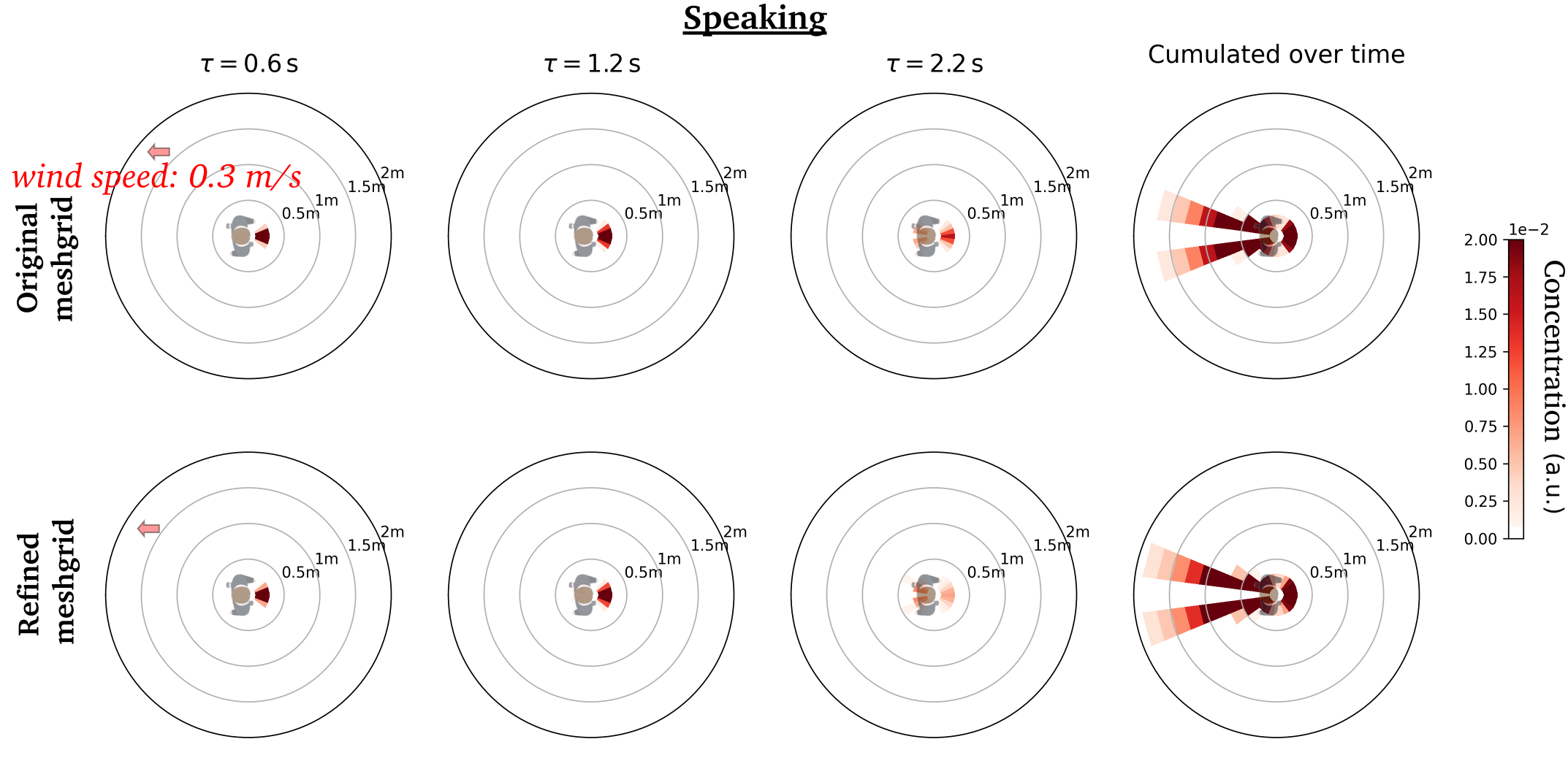}
    \caption{Comparison of the dynamic concentration maps generated at distinct time delays $\tau$ for a static pedestrian in a head wind blowing at $0.3\,\mps$, by coarse-graining CFD simulations performed with (\emph{top row}) the original meshgrid, (\emph{bottom row}) a refined meshgrid.}
    \label{fig:SI_FinerGrid_0p3ms}
    \end{center}
\end{figure*}

The grid is initially refined around the head of the manikin with a spatial resolution of 1~mm and coarsened further away. A dynamic mesh adaptation algorithm is used to refine the grid wherever needed. To do so, a passive scalar is injected at the mouth. Any location where the concentration of this passive scalar is non-zero is identified as a meaningful region and the grid is subsequently refined during the calculation, with a target grid size of 8~mm.  Figures~\ref{fig:SI_FinerGrid_0ms} and \ref{fig:SI_FinerGrid_0p3ms} prove that CFD
simulations with a finer grid size of 4~mm yield virtually identical coarse-grained concentration maps.

\subsection*{Boundary conditions}

The CFD database is parametrized by the incident air speed $v$ and the angle $\varphi$ between minus the incident velocity vector and the mouth direction (Fig.~\ref{fig:frames}).
The inflow boundary condition mimics an ambient wind, where a uniform flow of $\vctor v_w'= (-v \cos \varphi,v \sin \varphi,0)$ is imposed; the outflow boundary condition is applied on the other side of the domain, and slipping wall boundary conditions are applied to the lateral boundaries. 

The breathing flow is injected at the manikin's mouth, which was delimited by hand as the surface covered by the lips of the manikin, whose mouth is initially closed. This yields a non-planar surface of 4.7~cm$^2$ on which a uniform velocity is imposed, parallel to the ground and in front of the manikin. 

The time signal is periodic, with a period of~3.0 s, and was designed to mimic a breathing signal \cite{gupta2010characterizing} with a short period typical of the breathing pace while walking. Each breath is of 1~L of volume, so the breathing rate is 20 L$\cdot$min$^{-1}$.


\begin{figure}[ht]
    \begin{center}
	\includegraphics[width=0.6\textwidth]{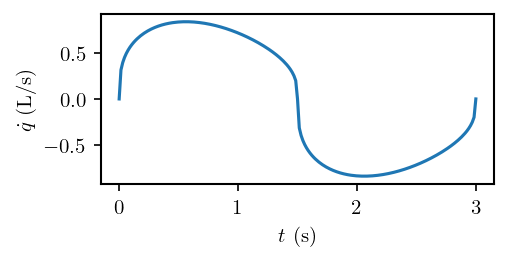}
    \caption{Flow rate imposed at the mouth of the manikin at each cycle of duration 3.0 s.}
    \label{fig:flow_rate}
    \end{center}
\end{figure}

\subsection*{Simulations}

Simulations are first performed over 4 cycles to install the flow. Then, 4 cycles are computed to collect the statistics presented in Fig.~\ref{fig:diagrams_wind} and the following maps. Solutions are stored every 0.25~s (12 par cycle) for statistical accumulation. 

\subsection*{Alternative physical modeling}
\label{sec:SI_alternativeCFD}
\begin{figure*}[ht]
    \begin{center}
	\includegraphics[width=\textwidth]{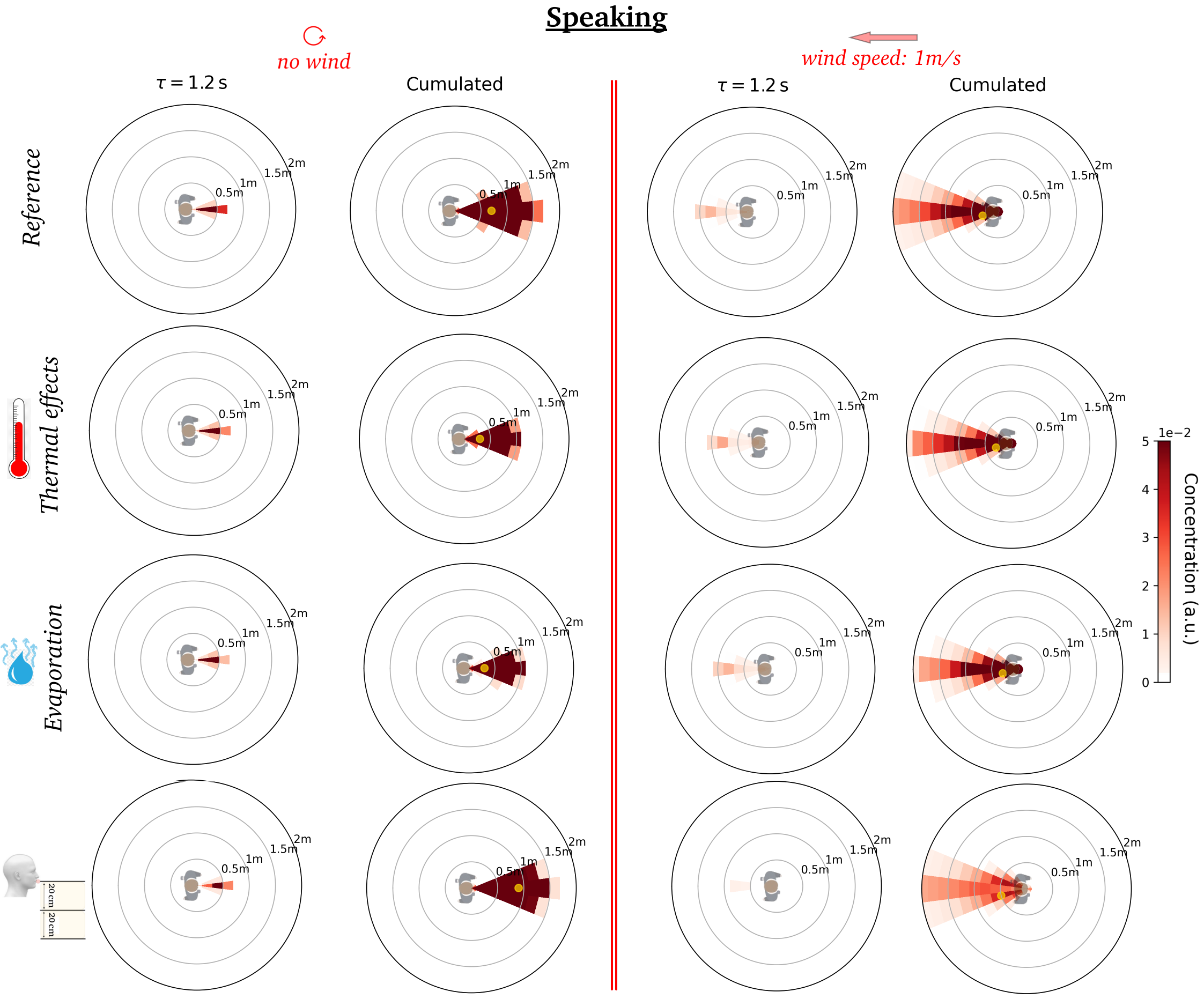}
    \caption{Comparison of the viral concentration maps obtained under different CFD modeling assumptions, for two wind speeds. The first row presents the reference simulations without thermal effects or evaporation in the puff; thermal effects are included in the second row, along with evaporation in the third one. Finally, the last row shows the result if the region of interest is shifted downwards by 20~cm, to compensate for the overlooked buoyancy effect due to the receiver's thermal plume.}
    \label{fig:SI_compDiagrams_CFD}
    \end{center}
\end{figure*}

The typical speeds of expiratory droplets and ventilation in conditions where there is a concern for contamination are well within the regime of incompressibility, i.e., low Mach numbers. However, thermal effects are present, which cannot be modeled with the incompressible Navier-Stokes equations presented in the previous sections. To test the influence of these effects, we performed a limited number of simulations with the variable density formulation of the Navier-Stokes equations applicable at low Mach number, as detailed in \cite{boulet2018modeling}. In these simulations, the air is represented as a mixture of nitrogen, oxygen, argon, carbon dioxide, and water, and its temperature  is set to $20^\circ C$, with a relative humidity of 50\%. The exhaled puff is warmer and more humid, with a temperature  set to $35^\circ C$ and a relative humidity of 90\%. The second row of Fig.~\ref{fig:SI_compDiagrams_CFD} illustrates the (moderate) impact of these thermal effects: In comparison with the incompressible model,incompressible model, droplets evade the region of interest somewhat closer to the point of emission, under windless conditions, because of buoyancy.

Next, the evaporation of droplets during their propagation was also introduced in the model. For that purpose, droplets were considered to be composed of pure water, until they shrunk to a diameter of one third of the original diameter, at which point evaporation was halted, to reflect the presence of a non-water content in the droplet \cite{Netz2020physics,Seyfert2022stability}. The model for evaporation follows the approach developed by Spalding \cite{spalding_combustion_1953}, similar to the one described by Bale \textit{et al.} \cite{Bale2022quantifying}. The impact of evaporation on the concentration maps, as compared to those with thermal effects only, is barely noticeable (Fig.~\ref{fig:SI_compDiagrams_CFD}).
It should be noted that finer models for the evaporation of droplets of respiratory fluids (instead of water) are available in the literature \cite{Seyfert2022stability}, but were not implemented in our tests. They would certainly further restrict the effect of evaporation by slowing down the decrease of diameter in time.  

Finally, the thermal plume due to the receiver's body heat may lift droplets upwards and was claimed to facilitate the airborne transmission of the virus in enclosed spaces \cite{sun2021human}. Our dynamic concentration maps are oblivious to the presence of a receiver, but we investigated the possible impact of these ascending flows by shifting the region of interest downwards by 20~cm (to compensate for these potentially overlooked ascending flows). Figure~\ref{fig:SI_compDiagrams_CFD}(bottom row) ascertains that this shift only has a moderate effect on the concentration maps.

So far, the effects of these physical details of the CFD simulations were assessed in windless conditions and found to be modest. As soon as an external wind is added, these effects become slighter [Fig.~\ref{fig:SI_compDiagrams_CFD}(right)], as the droplets are swept away by the wind.

\begin{figure*}[ht]
    \begin{center}
	\includegraphics[width=\textwidth]{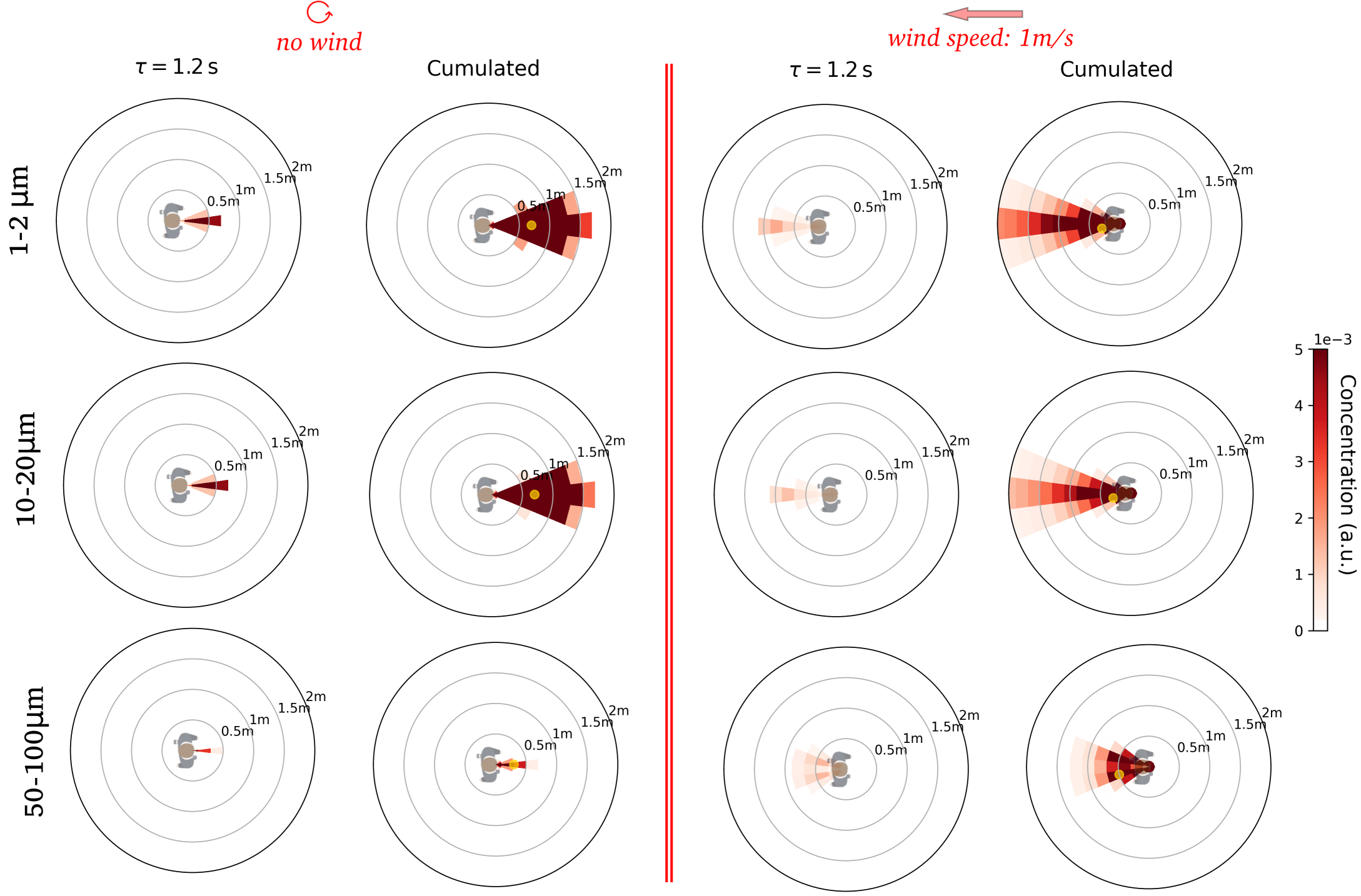}
    \caption{Comparison of the viral concentration maps corresponding to different ranges of droplet sizes, for two wind speeds.}
    \label{fig:sizeDiagrams_CFD}
    \end{center}
\end{figure*}

\section*{Transmission via large droplets}   
\label{sec:SI_large_droplets}
\subsection*{Inhalation of large droplets}

In the main text, the focus was put on micron-sized respiratory droplets, which are much more easily inhaled than their larger counterparts ($\geqslant 50\,\mathrm{\mu m}$ after evaporation) originating from the oral cavity and lips \cite{johnson2011modality}.

\begin{figure}[ht]
    \begin{center}
	\includegraphics[width=0.9\columnwidth]{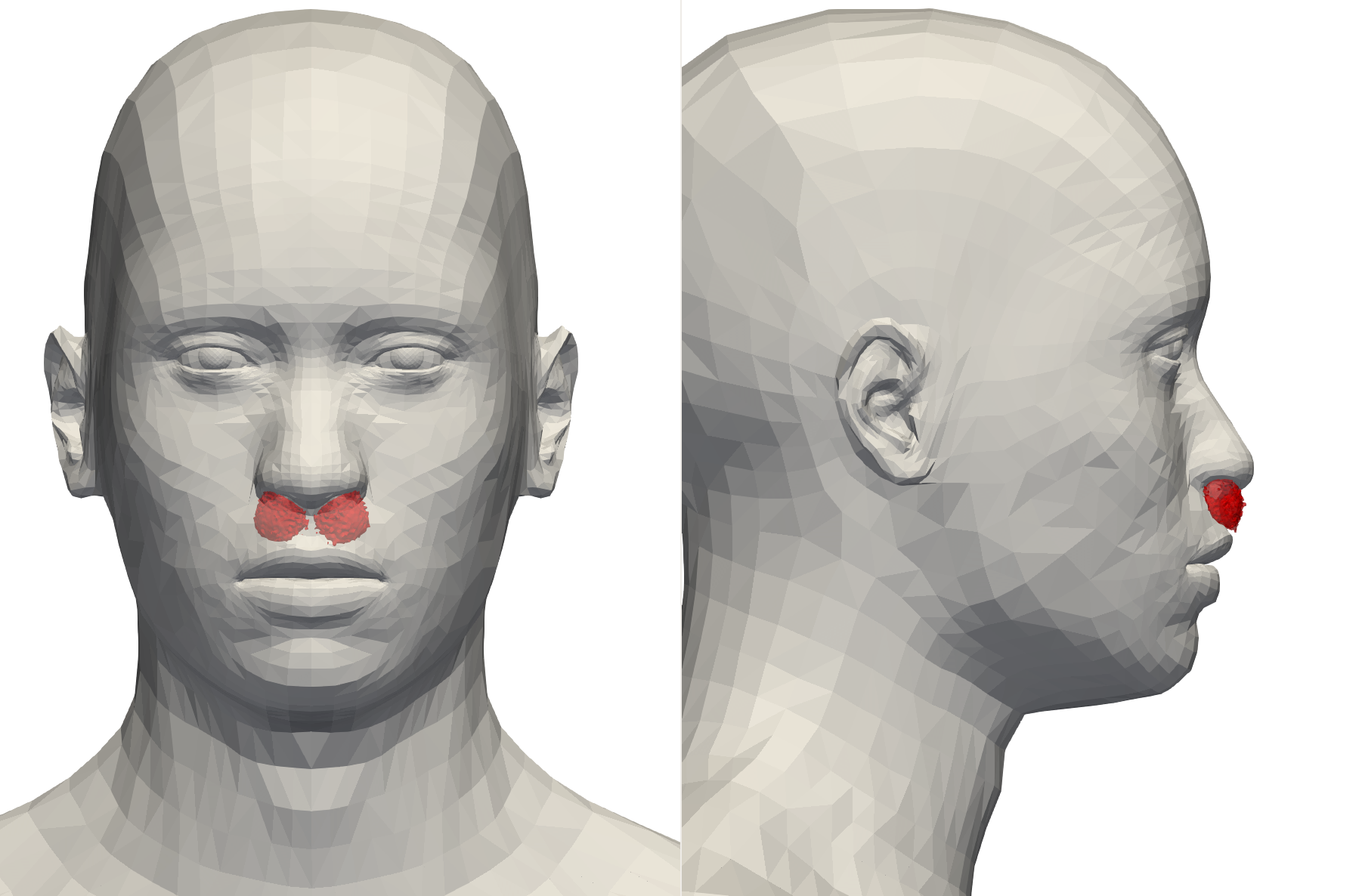}
    \caption{Region of the simulation domain where the vertical air flow during inhalation reaches a speed of at least $0.3\,\mps$, as required to compensate the downward sedimentation of  droplets of diameter about $100\,\mathrm{\mu m}$. The inhaling flow rate is $0.55 \mathrm{L\,s^{-1}}$.}
    \label{fig:isospeed}
    \end{center}
\end{figure}

Indeed, an inhalation flow speed of around  $0.3\,\mps$ in the vertical direction is needed to overcome the sedimentation of a $\sim 100\,\mathrm{\mu m}$-droplet. 
A dedicated simulation has been performed to delineate the region in space where the vertical flow velocity exceeds $0.3\,\mps$ during inhalation. A flow rate of $0.55 \mathrm{L\,s^{-1}}$ is imposed, which is a rather high value for calm breathing \cite{gupta2010characterizing}.  Even so,
Fig.~\ref{fig:isospeed} shows that such a vertical speed of $0.3\,\mps$ is only reached extremely close to the nostrils, which makes them less plausible candidates for inhalation than smaller droplets.

\subsection*{Dynamic maps of concentration}
\begin{figure*}[ht]
    \begin{center}
	\includegraphics[width=1\textwidth]{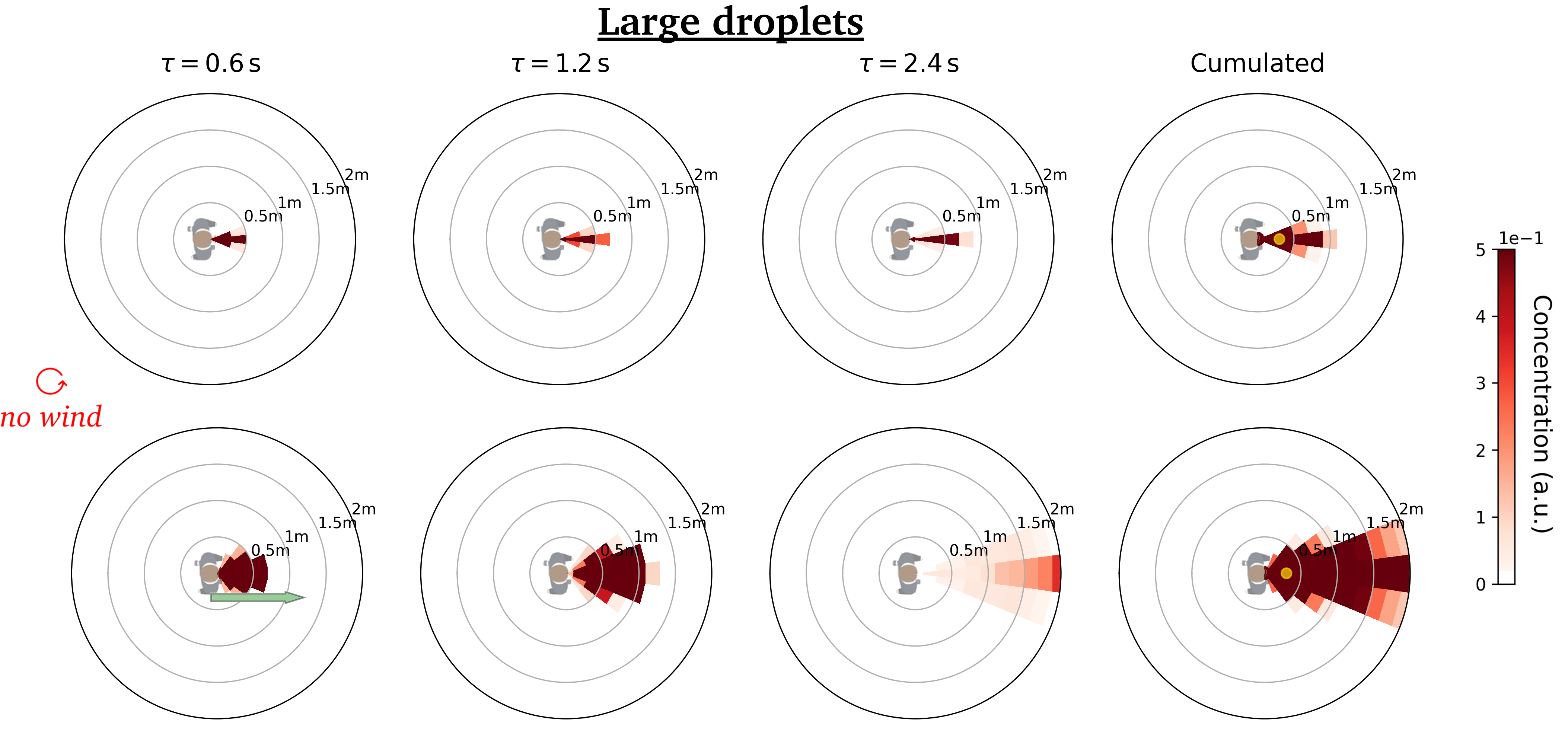}
    \caption{Dynamic concentration maps associated with the oral mode (large droplets) emitted while speaking, in windless conditions. (Top row) static emitter, (bottom row) walking emitter. }
    \label{fig:SI_maps_LargeDroplets}
    \end{center}
\end{figure*}

Nevertheless, we computed the dynamic maps of viral concentration associated with the emission of these large droplets, the so-called oral mode; they are shown in Fig.~\ref{fig:SI_maps_LargeDroplets}. Clearly, these maps differ from those obtained for smaller aerosols: In windless conditions, they are transported over less than one meter before sedimenting. The walking motion drags these droplets farther ahead, whereas the external wind impacts them diversely: it hardly affects the mostly ballistic motion of the heaviest ones, while others are carried away by the wind. 
It follows that the interpolation method proposed in Sec.~\ref{sub:interpolation} does not work
quite as well on this broad class of droplet sizes, but tests show that this interpolation still
yields reasonable concentration maps; it will thus be retained.

\subsection*{Rates of new infections via large droplets}

The inclusion of the oral mode substantially alters the ranking of transmission risks across the scenarios under study, with street caf\'es
that become outshadowed by the observed outdoor market. This can be ascribed to the short range of propagation of the large droplets when they are emitted by a static person (see Fig.~\ref{fig:SI_maps_LargeDroplets}), as compared to a moving pedestrian.
In parallel, owing to their larger sizes, droplets follow a more ballistic trajectory and are less sensitive to the wind, as reflected by the limited effect of the external wind on the rate of new infections in the different scenarios.

\begin{figure*}[ht]
    \begin{center}
	\includegraphics[width=1\textwidth]{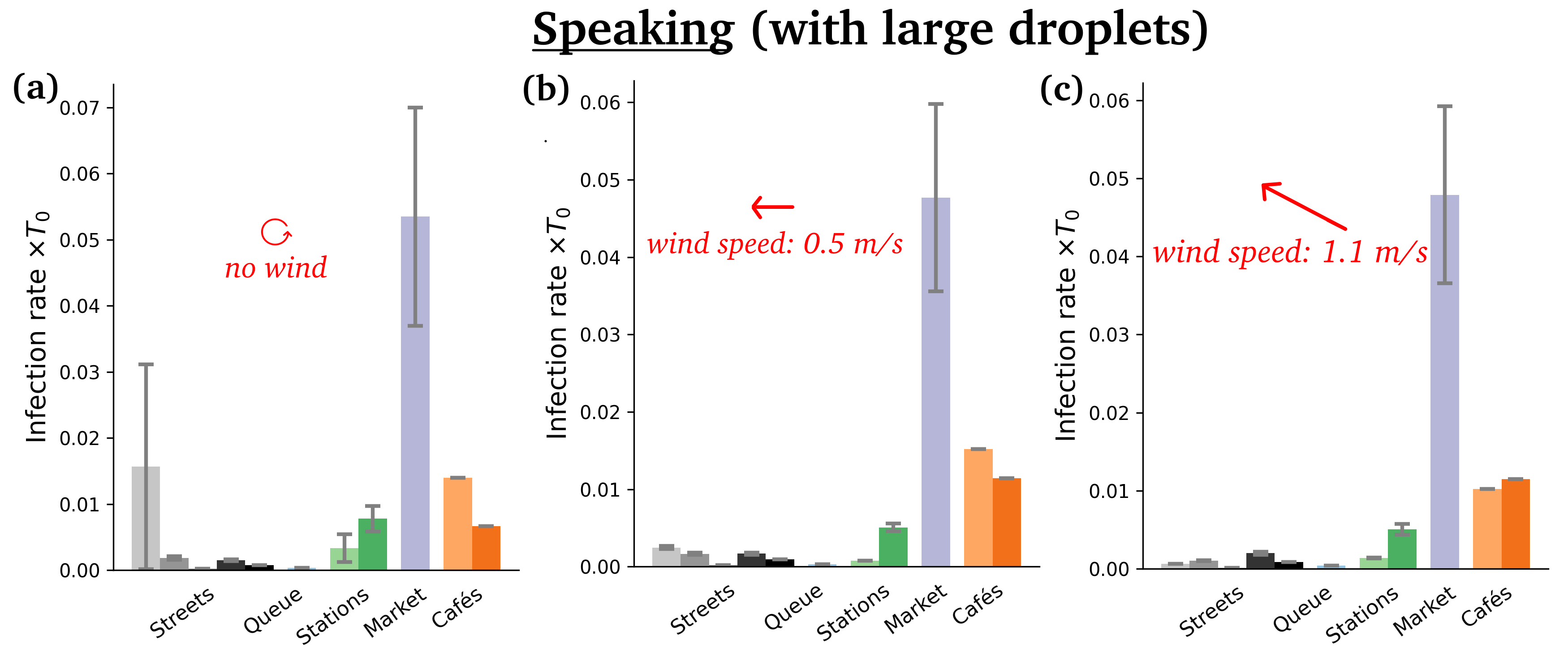}
    \caption{Transmission risks associated with the oral mode (large droplets, assumed to be inhalable here) emitted while speaking, in the different scenarios under study.}
    \label{fig:histo_LargeDroplets}
    \end{center}
\end{figure*}

\section{Additional figures}

\begin{figure*}[ht]
    \begin{center}
	\includegraphics[width=1\textwidth]{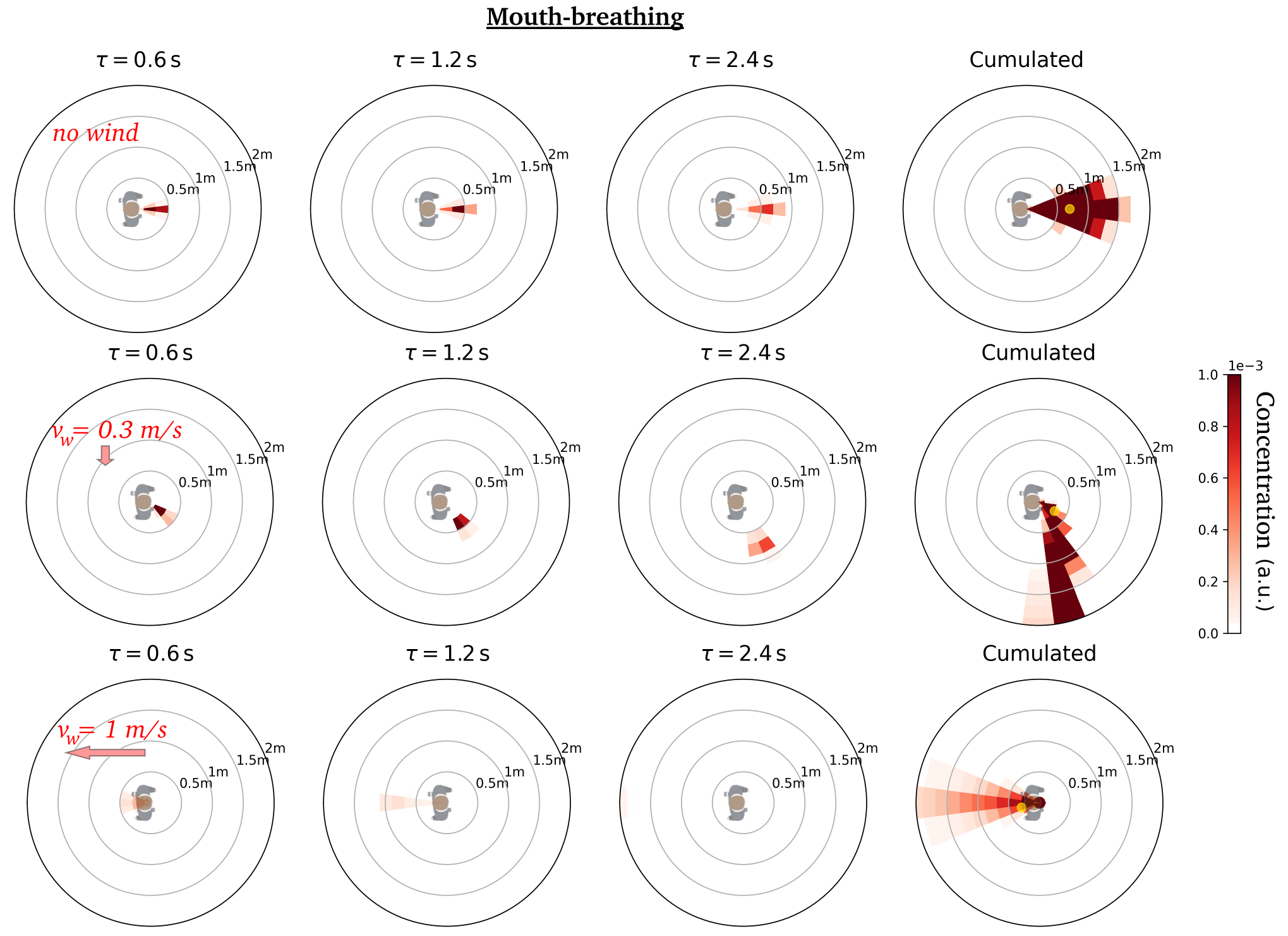}
    \caption{Dynamic maps of viral concentration associated with breathing through the mouth : Effect of the external wind. (Top row) no wind, (middle row) lateral wind at $0.3\,\mps$, (bottom row) head wind at $1.0\,\mps$.  Note the very different scale, as compared to the maps associated with speaking.}
    \label{fig:SI_diagrams_breathing_wind}
    \end{center}
\end{figure*}

\begin{figure*}[ht]
    \begin{center}
	\includegraphics[width=1.0\textwidth]{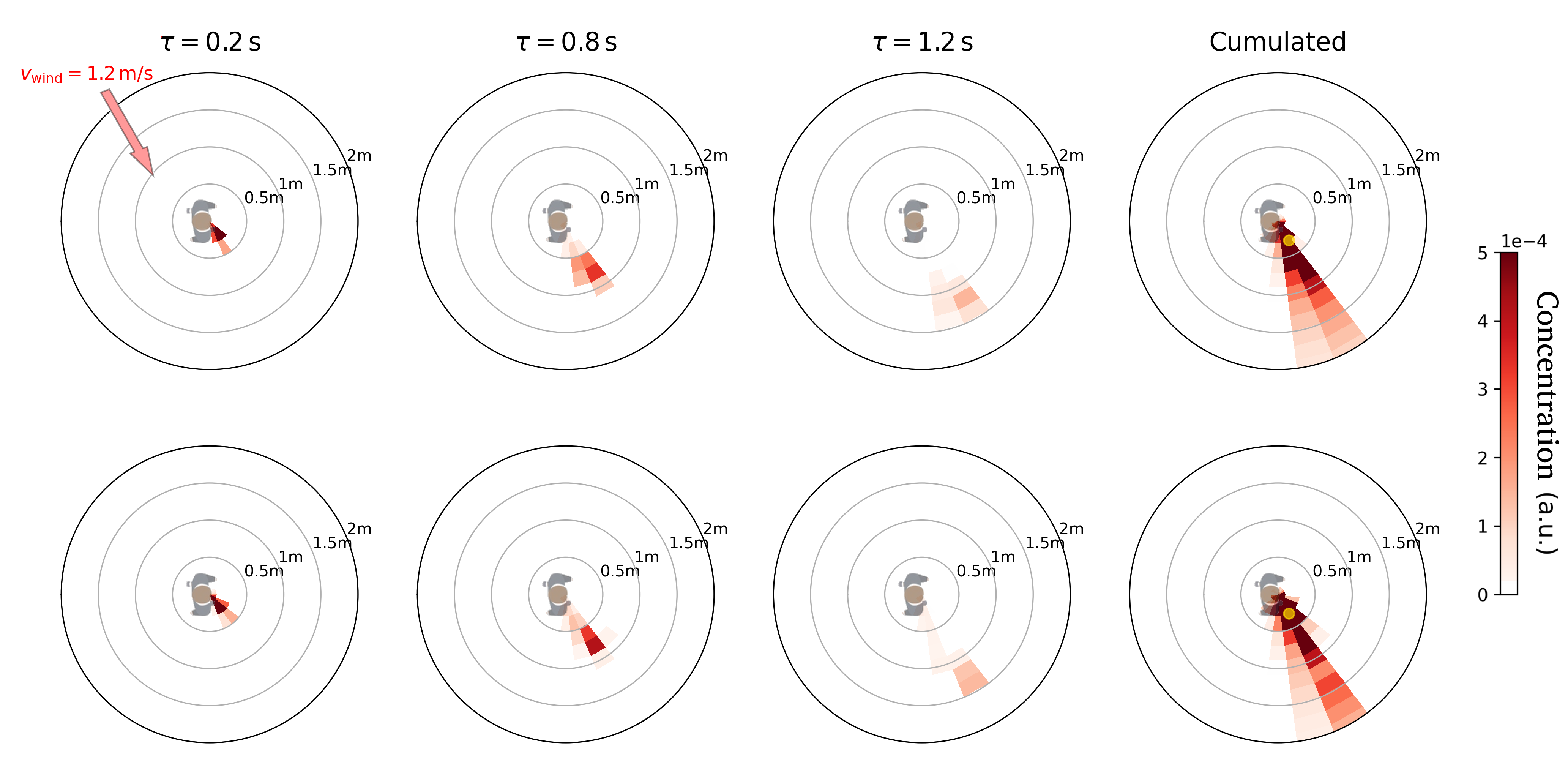}
    \caption{Interpolation of dynamic concentration maps at for $v=1.2\,\mps$ and $\varphi=115^{\circ}$. (Top) Coarse-grained map calculated using a \emph{bona fide} CFD simulation corresponding to these specific conditions. (Bottom) Concentration map obtained by interpolation.}
    \label{fig:SI_diagrams_interp}
    \end{center}
\end{figure*}

\begin{figure*}[ht]
    \begin{center}
	\includegraphics[width=1\textwidth]{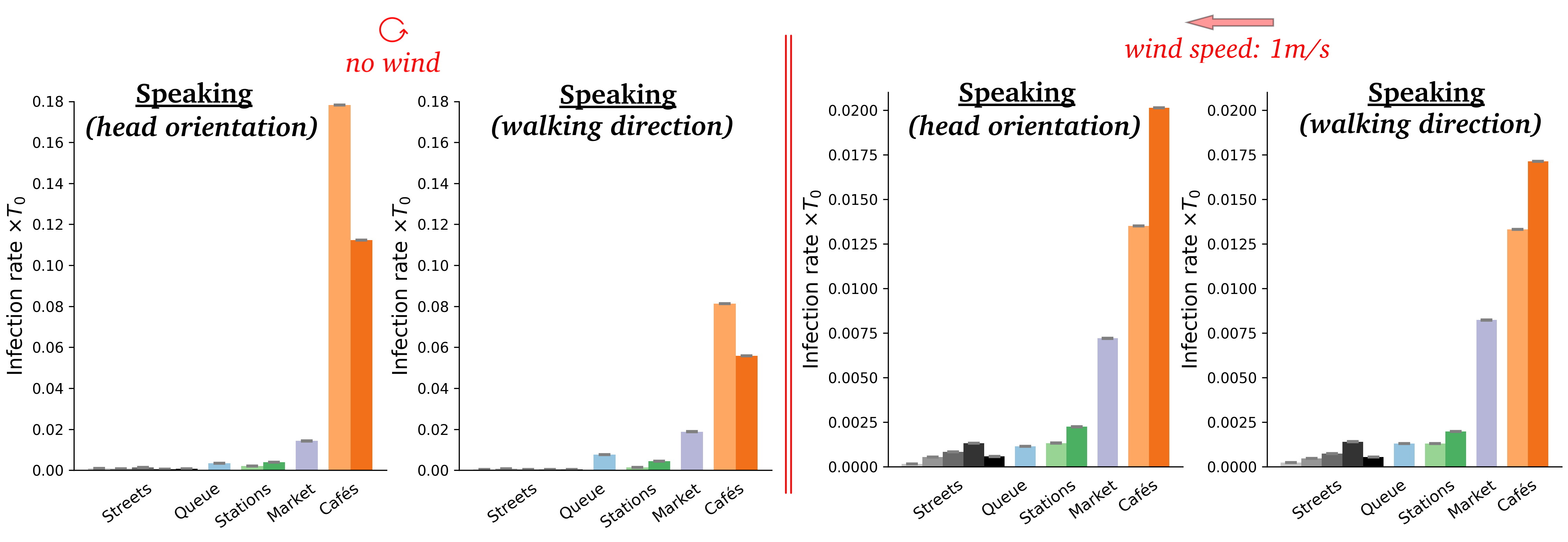}
    \caption{Variation of the transmission risks associated with speaking in the different scenarios under study for two distinct wind conditions, depending on the modeling assumptions regarding emission: Either (i) the head orientation governs the directions of emission and walking, or (ii) the direction of emission is supposed to align with the walking direction.  The estimated risks are fairly similar under both assumptions, except in scenarios where many people are at a halt (street caf\'es, waiting line) and the walking direction in (ii) is thus ill-defined.}
    \label{fig:SI_histo_WalkingDir}
    \end{center}
\end{figure*}

\end{document}